\documentclass[twocolumn]{aastex62}
\newcommand{\be}{\begin{equation}}
\newcommand{\ee}{\end{equation}}
\newcommand{\ba}{\begin{eqnarray}}
\newcommand{\ea}{\end{eqnarray}}
\newcommand{\bnabla}{\mbox{\boldmath$\nabla$}}

\newcommand{\nn}{\mbox{} \nonumber \\ \mbox{}}
\newcommand{\BQ}{B_{\rm Q}}
\newcommand{\alphem}{\alpha_{\rm em}}
\usepackage{graphicx}
\usepackage{epstopdf}
\usepackage{mathtools}

\begin{document}

\title{Pair Plasma in Super-QED Magnetic Fields \\ and the Hard X-ray/Optical Emission of Magnetars}

\author{Christopher Thompson}
\affiliation{Canadian Institute for Theoretical Astrophysics, 60 St. George Street, Toronto, ON M5S 3H8, Canada}  
\author{Alexander Kostenko}
\affiliation{Department of Astronomy and Astrophysics, University of Toronto, 50 St. George Street, Toronto, ON M5S 3H4, Canada}

\date{\today}

\begin{abstract}
  The photon spectrum emitted by a transrelativistic pair plasma is calculated in the presence of
  an ultrastrong magnetic field, and is shown to bear a remarkable resemblance to the rising hard X-ray spectra
  of quiescent magnetars.  This emission is powered by pair annihilation which, in contrast
  with a weakly magnetized pair plasma, shows an extended low-frequency tail similar to bremsstrahlung.
  Cross sections for electron-positron annihilation/scattering, two-photon pair creation, and photon-$e^\pm$ scattering are
  adopted from our earlier ab initio QED calculations in the regime $10\alphem^{-1}\BQ \gg
  B \gg \BQ$.   Careful attention is given to the $u$-channel scattering resonance.
  Magnetospheric arcades anchored in zones of intense crustal shear and reaching about twice the magnetar radius
  are identified as the sites of the
  persistent hard X-ray emission.  We deduce a novel and stable configuration for the magnetospheric circuit, with a high
  plasma density sustained by ohmic heating and in situ pair creation.  Pairs are sourced non-locally by photon collisions
  in zones with weak currents, such as the  polar cap.  Annihilation bremsstrahlung extends to the optical-IR band,
  where the plasma cutoff is located.  The upper magnetar atmosphere experiences strong current-driven growth of
  ion-acoustic turbulence, which may limit positron diffusion.  Coherent optical-IR emission is bounded near the observed flux by induced scattering.
  This model accommodates the rapid X-ray brightening of an activating magnetar,
  concentrated thermal hotspots, and the subdominant thermal X-ray emission
  of some active magnetars.  Current injection is ascribed to continuous magnetic braiding, as seen in
  the global yielding calculations of Thompson, Yang \& Ortiz.
\end{abstract}


\section{Introduction}

Extensive observational and theoretical studies of magnetars over the last two
decades have made clear that magnetic field decay is mediated not only by
internal diffusive processes, but also by the excitation and damping
of powerful electric currents outside the star.  The output of magnetars in hard X-rays,
even in relative quiescence, can outstrip the spindown power
by 3-4 orders of magnitude \citep{kuiper04,kuiper06,mereghetti05a,gotz06};  the output in the optical-IR
band can exceed the expected surface Rayleigh-Jeans emission by
an even larger factor \citep{hulleman01,hulleman04,durant05}.  The emission in these two bands is sometimes correlated
in a striking way \citep{tam04}.  \cite{KB17} give an up-to-date review of the rich observational developments
over the last two decades and recent theoretical approaches.

Only fragmentary progress has been
made toward understanding how such powerful currents could be sustained, given the
strong gravitational stratification of the magnetar atmosphere.  It is clear
that a self-consistent model of the magnetospheric plasma, which incorporates electron-positron pair creation,
is a pre-requisite for identifying the source of the non-thermal emission.  The problems of photon creation
and pair creation must be tackled together.

Our focus here is on the simplest possible plasma state: a collisional, transrelativistic,
and quasi-thermal pair gas.   We show that a single quantum electrodynamic (QED)
process operating in such a plasma
can easily account for the measured hard X-ray continuum
of a quiescent magnetar (which emits an increasing energy flux above $\sim 10$ keV in photon energy).  This process, which we
term ``annihilation bremsstrahlung'', operates only in magnetic fields exceeding
$\BQ \equiv m_ec^3/e\hbar = 4.4\times 10^{13}$ G.  It generates an extended
electromagnetic continuum in the absence of a relativistic distribution
of particle energies, and without the need to invoke multiple scattering.
The source of the $> 10$ keV emission of a quiescent magnetar is found to be
concentrated in the inner magnetosphere, where the magnetic flux density is around
$5\,\BQ = 2\times 10^{14}$ G, somewhat weaker than the surface field.

The emission of a soft photon (a photon with energy much less than the electron rest mass, $m_ec^2$)
occurs with a dramatically enhanced cross section during the collision of an electron and positron,
if the pair temporarily converts to a gamma ray, $e^+ + e^- \rightarrow \gamma \rightarrow e^+ + e^-$.
The emission rate in this annihilation
channel ($s$ channel) is enhanced by a factor $10^2$ or more as compared with the scattering
channel ($t$ channel), which closely resembles classical bremsstrahlung.  The high collision rate
within the plasma  also implies
strong ohmic heating.  In order to radiate the dissipated energy, we show that the pair density must
exceed, by at least an order of magnitude, the minimum density that will supply the
magnetospheric current.  Above a threshold current density, the plasma naturally falls into a collisional and trans-relativistic state,
very different from the relativistic double layer uncovered by \cite{BT07}.

Our simulations employ the rates and cross sections of several QED processes that were obtained in the
ab initio calculations of \cite{KT18,KT19}.   We develop a detailed Monte Carlo
description of photon emission by $e^\pm$ annihilation, absorption by photon collisions,
and redistribution by multiple scattering, taking into account the strong $u$-channel resonance experienced
by photons near the pair creation threshold.  Not only does this procedure yield self-consistent
photon spectra, but it allows us to track the balance of pair creation and annihilation.
We further show that the trans-relativistic, collisional plasma is an attractor, in the sense
that the plasma will spontaneously move toward a state of combined energy and annihilation equilibrium.

The hard X-rays are emitted almost entirely in the ordinary polarization mode (O-mode), which has
a much larger cross section for interaction with $e^\pm$ in an ultrastrong magnetic field than
does the extraordinary mode (E-mode; \citealt{HL06}).  In fact, the presence of a rising hard X-ray 
continuum in the magnetar energy spectrum can be viewed, in our approach, as partly a consequence of this strong
imbalance.  Heat deposited below the magnetar surface by crustal yielding or impacting relativistic
particles is mostly radiated through the low-cross section E-mode.   The closest analog amongst previous
models of the hard X-ray emission is the bremsstrahlung model described by \cite{TB05}, which involves a warm beam-heated layer
in the upper magnetar atmosphere.  Classical bremsstrahlung emission decays away from the magnetar surface much
more rapidly than does the annihilation bremsstrahlung process described here.  Calculations combining the magnetospheric
emission with surface heating by annihilating positrons will be presented in a companion paper \citep{KT20}.

The current-carrying flux structures that support the collisional pair plasma described here are associated with
zones of intense crustal shear.  These are produced when the core magnetic field comes into imbalance
and globally stresses the magnetar crust \citep{TYO17}, or global Hall drift creates strong localized magnetic shear
\citep{gourg16}.  The surface blackbody emission from impacting charges is relatively weak as compared
with the relativistic double layer solution of \cite{BT07}, allowing for relatively stronger output
in the hard X-ray band.

The emitted hard X-ray spectrum is predicted to extend to an energy $\hbar\omega \sim m_ec^2$.  As a result,
collisions between photons of this peak energy are a powerful source of pairs outside the parts of
the magnetosphere that connect to the crustal shear zones.  In many cases, the crustal shear zones will not
intersect with the polar cap region.  Our electrodynamic model therefore suggests a significantly weaker pair density in the
outer magnetosphere as compared with the ``j-bundle'' construction of \cite{B09}.

The same current-carrying structures that produce the hard X-ray continuum of a magnetar are also a promising source
of the bright optical-IR emission.   Annihilation bremsstrahlung extends down into the optical-IR band, but is cut off
below $\sim 10^{14}$ Hz by a combination of the finite plasma frequency, self-absorption, and induced scattering.
This process is not relevant to the bright detected
radio to millimeter-wave emission \citep{camilo06,camilo07,torne15,torne20}.  Although the observed optical-IR output
is too bright to be produced directly by annihilation bremsstrahlung, it could represent reprocessing of intrinsically
brighter UV emission.  Given that the peak plasma frequency sits in the optical-IR band, the transrelativistic pair plasma
is also a promising source of coherent radiation \citep{eichler02}.  The IR flux as limited by induced $e^\pm$ scattering
turns out to be similar to the observed flux.

The plan of this paper is as follows.  Section \ref{s:int} describes the relevant QED interactions between
photons, electrons and positrons, in particular the two-photon annihilation spectrum of transrelativistic
pairs and how this varies with the background magnetic field.  Section \ref{s:monte} explains the MC method.
The equilibrium state of an $e^\pm$ plasma and the associated ohmic heating are investigated in Section \ref{s:current},
along with the current-driven instability of the magnetar atmosphere.
Analytic estimates and detailed MC results
for the emergent photon spectrum and rates of pair creation/destruction are presented
in Section \ref{s:results}.  Optical/IR plasma emission from
the same plasma is considered in Section \ref{s:optical}, along with the relevant physical constraints.
A concrete example of a current-carrying arcade is presented in Section \ref{s:arcade}, and 
the calculated non-thermal emission is briefly compared with data.
A summary is presented in Section \ref{s:summary}.
There we also compare our approach with the previous resonant scattering models of \cite{FT07}, \cite{BH07}, \cite{nobili08a} and \cite{B13a},
and examine the implications of gamma-ray emission in the inner magnetosphere for the global current flow, and for
the decay of magnetospheric currents.
The Appendix offers details of soft-photon emission in $e^\pm$ backscattering.

Throughout this paper, we adopt the shorthand $X = X_n\times 10^n$ to describe the normalization of quantity $X$
in c.g.s. units.

\section{Electron-Positron-Photon Interactions}\label{s:int}

The interactions of electrons, positrons, and photons are modified in some remarkable ways by an ultrastrong magnetic
field, exceeding $\BQ$.  Our focus here is on the emission and transport of ordinary-mode (O-mode) photons in a plasma
containing mildly relativistic electrons and positrons.  The pairs remain confined to the lowest Landau state, and
the plasma is dilute enough that the dielectric tensor is dominated by vacuum polarization.
The electric vector of the O-mode overlaps with the background magnetic field ${\bf B}$
($\hat E \parallel \hat k\times(\hat k\times {\bf B})$ given a propagation direction $\hat k$).
Its low-frequency (classical) scattering cross section\footnote{The QED rates and cross sections presented here only
  involve the component of the photon polarization vector parallel to $\hat B$, meaning that the $(\omega m_ec/eB)^2\sigma_T$ term in the classical
photon-electron  cross section is effectively neglected.  For gamma rays with energies of order $(1-2)m_ec^2$, the resonant enhancement
  in the scattering cross section emerging from the QED calculation is much the stronger effect.}
is $\sin^2\theta\,\sigma_{\rm T}$, where $\theta = \cos^{-1}(\hat k\cdot {\bf B}/B)$ \citep{HL06}.

While enhanced scattering at Landau resonances has received much attention \citep{DH86,gonthier00,nobili08b,mushtukov16}, 
the interactions of $e^\pm$ that are confined to the lowest Landau state are of great interest in the magnetar
regime, $10\alphem^{-1}\BQ \gg B \gg \BQ$ (where $\alpha_{\rm em} \simeq 1/137$ is the fine structure constant).
Not only do these interactions simplify
dramatically in this regime  -- vacuum polarization corrections remaining weak --
but they exhibit a number of novel features \citep{KT18, KT19}:

\begin{figure}
  \epsscale{1.2}
  \vskip -0.3in
\plotone{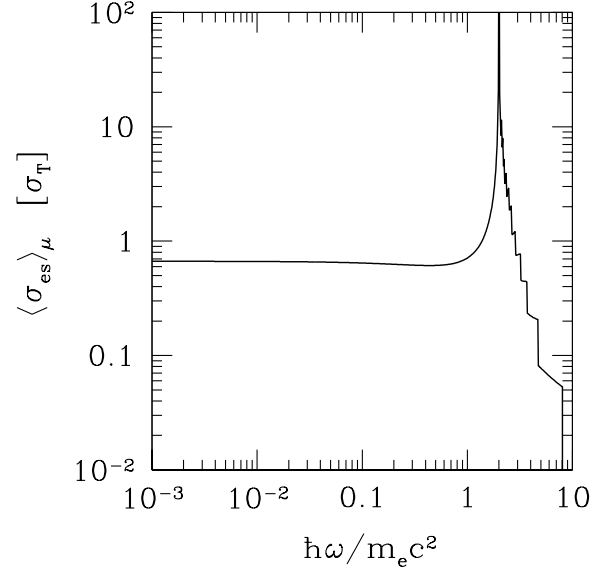}
\vskip -0.8in
\caption{Angle-averaged cross section for the scattering of photons off cold electrons. For a given photon
  energy $\hbar\omega$, the average only includes directions below threshold for pair conversion.  Structure
  at $\hbar\omega > 2m_ec^2$ reflects the direction cosine binning employed in the MC code ($N_\mu = 2^5$ in this case).}
\end{figure}\label{fig:tauscatt}

\begin{figure}
  \epsscale{0.9}
  \vskip -0.3in
\plotone{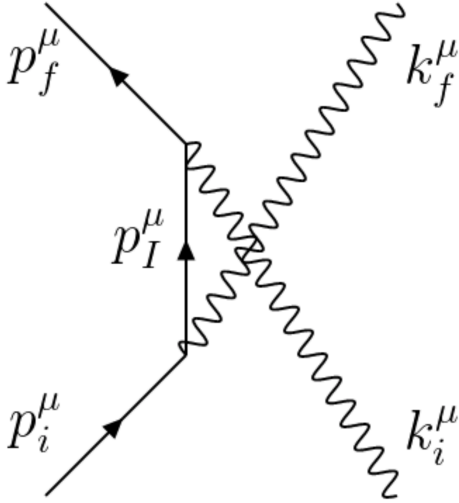}
\caption{Feynman diagram contributing to the $u$-channel resonance between the initial-state electron
  (taken to be at rest in the lowest Landau state, $p^z_i = 0$) and the final-state photon (wave four-vector $k^\mu_f$).}
\end{figure}\label{fig:uchannel}

\begin{figure*}
  \epsscale{1.1}
  \vskip -0.3in
\plottwo{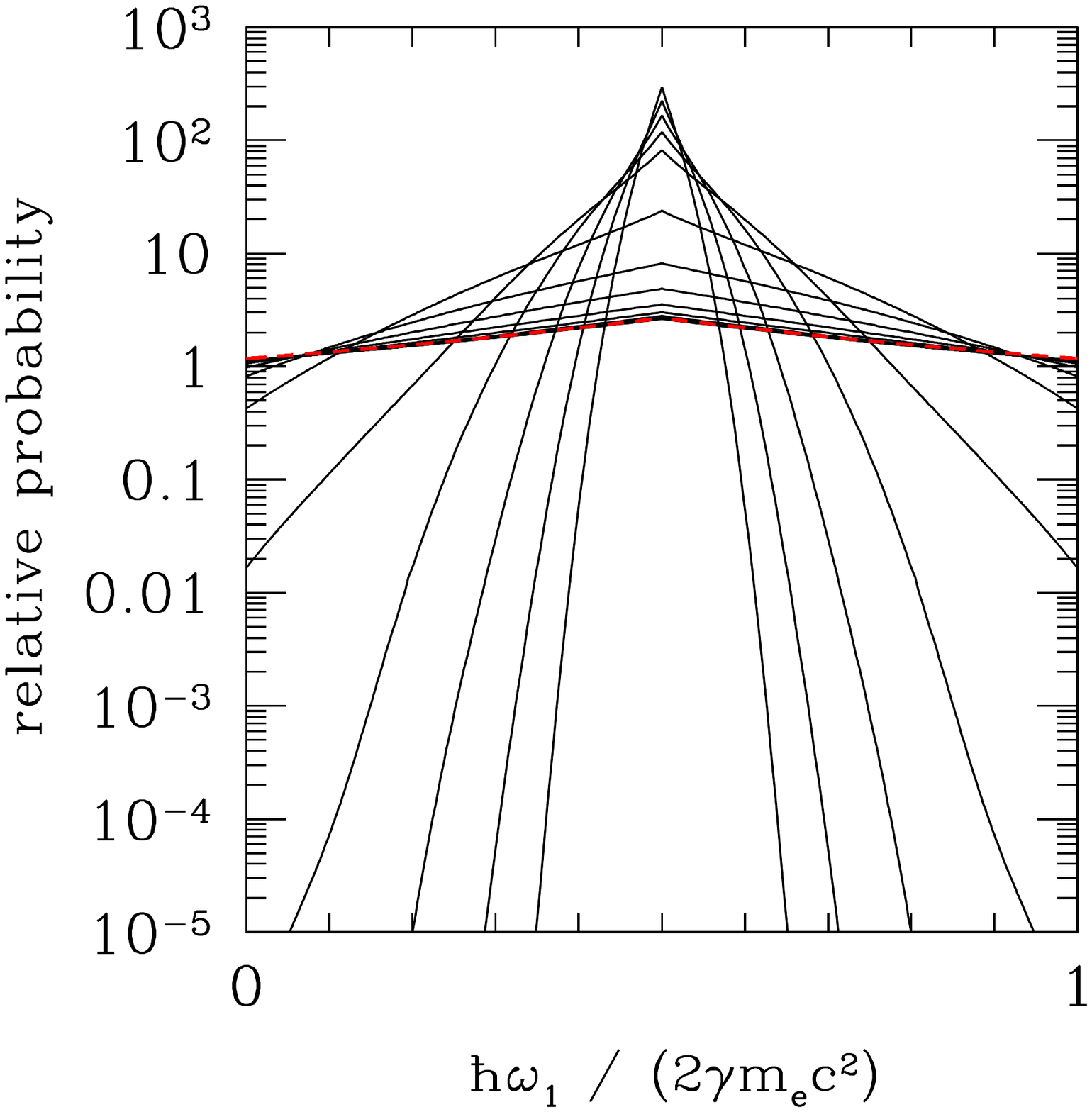}{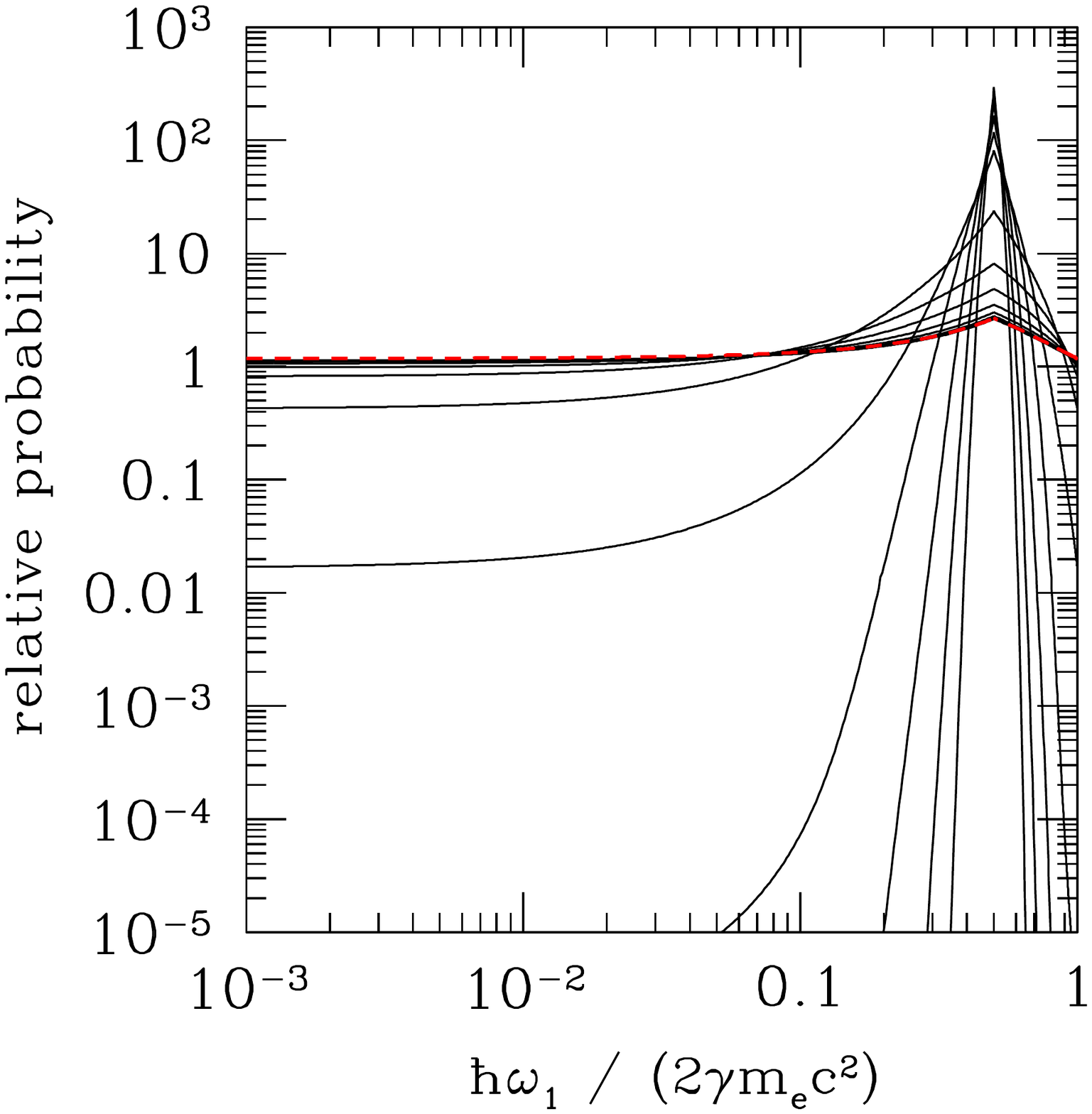}
\vskip -0.8in
\caption{Relative probability that photons of energy $\hbar\omega_1$ and $2\gamma m_ec^2 -\hbar\omega_1$
  are emitted in the annihilation of a $e^\pm$ pair, which is confined to the lowest Landau level in a strong magnetic field.
  (Left panel:  linear frequency axis.   Right panel:  logarithmic, showing the low-frequency tail.)
   Red dashed curves:  high-$B$ approximation of \cite{KT18}.
  Black curves show a range of flux densities extending from $10^{-1.6}\,\BQ = 1.1\times 10^{12}$ G (most peaked)
  to  $10^2\,\BQ$ (least peaked), separated by $0.3$ in $\log_{10}B$.  Here $\gamma = \sqrt{2}$ in the center-of-momentum frame.
  Black curves are obtained by an extended sum over the intermediate-state Landau level,
using the matrix elements of \cite{DB80}.}
\end{figure*}\label{fig:omega_ratio}

(i) $e^\pm + \gamma \rightarrow e^\pm + \gamma$.  Electron-photon scattering shows a strong $u$-channel resonance when the photon energy approaches the threshold
for single-photon pair creation, $\hbar\omega \rightarrow 2m_ec^2/\sin\theta$
(Figure \ref{fig:tauscatt}).  Very close to this resonance the scattered photon can 
convert directly to a magnetized pair.
In other words, the cross section for scattering-induced pair creation,
$\gamma + e^\pm \rightarrow \gamma + e^\pm \rightarrow e^+ + e^- + e^\pm$ is
strongly enhanced (by a factor $\alphem^{-1}$ in addition to the resonance effect) compared with an unmagnetized vacuum
\citep{KT18}.

  The resonance involves a vertex between the initial-state electron and the final-state photon (Figure \ref{fig:uchannel}).
  The resonance arises when the intermediate-state electron approaches the mass shell.
  The intermediate-state electron four-momentum $p^\mu_I$ is related to those of
  the initial-state electron ($E_i = m_ec^2$, $p^z_i = 0$) and final-state photon ($\hbar\omega_f$, $\hbar k^z_fc = \mu_f \hbar\omega_f$)
  by $p^\mu_I = p^\mu_i - \hbar k^\mu_f c$.  The mass-shell condition $p^\mu_Ip^{\mu I}c^2 = E_I^2 - (p^z_I)^2c^2 = (m_ec^2)^2$
  corresponds to $\hbar\omega_f = 2m_ec^2/(1-\mu_f^2)$, which is above threshold for single-photon pair conversion. This resonance
  appears in the case of magnetic $e^\pm$-photon scattering (it is not present in non-magnetic Klein-Nishina scattering) because
  momentum conservation transverse to ${\bf B}$ involves the term $\pm e{\bf A}_\perp/c$
  in the generalized momentum of the scattering $e^\pm$, where ${\bf A}_\perp$ is the background vector potential.

(ii) $e^+ + e^- \rightarrow \gamma + \gamma$.
A significant flux of these near-threshold gamma-rays is produced by the annihilation of pairs into two photons.
There is a strong contrast in the annihilation process between magnetic fields that are weaker versus stronger than $\BQ$.
Figure \ref{fig:omega_ratio} shows that in magnetic fields that are characteristic of ordinary pulsars ($10^{12}$ G)
the annihilation energy is approximately equally divided between the two photons, whereas the energy is
broadly distributed in frequency in magnetic fields stronger than $\sim 5\,\BQ$.  

More generally, two-photon $e^+-e^-$ annihilation divides into three distinct channels, depending on
whether two, one, or no photon exceeds the threshold for pair conversion:
\vskip .05in
1.  $e^+ + e^- \;\rightarrow\; \gamma + \gamma \;\rightarrow\; e^+ + e^- + e^+ + e^-$ \quad\quad (2-p);

2.  $e^+ + e^- \;\rightarrow\; \gamma + \gamma \;\rightarrow\; \gamma + e^+ + e^-$ \quad\quad (1-$\gamma$/1-p);

3.  $e^+ + e^- \;\rightarrow\; \gamma + \gamma$ \quad\quad (2-$\gamma$).
\vskip .05in
\noindent The emission of two pair-converting photons is kinematically allowed
when $\gamma > 2$ in the center-of-momentum frame of the colliding pair.
Two-photon annihilation is a sink for pairs when they are cool (sub-relativistic),
and a source when they are relativistically warm.

The partial cross sections of these three channels are compared in Figure \ref{fig:sigma_ann};  they
are obtained by integrating Equation (\ref{eq:sigma_ann_m1m2}) below over the relevant parts of the photon phase space.
As a result of a soft-photon divergence, the cross section of the 1-$\gamma$/1-p channel dominates that of
the 2-$\gamma$ channel, except at low collision speeds.

\begin{figure}
  \vskip -0.3in
\epsscale{1.2}
\plotone{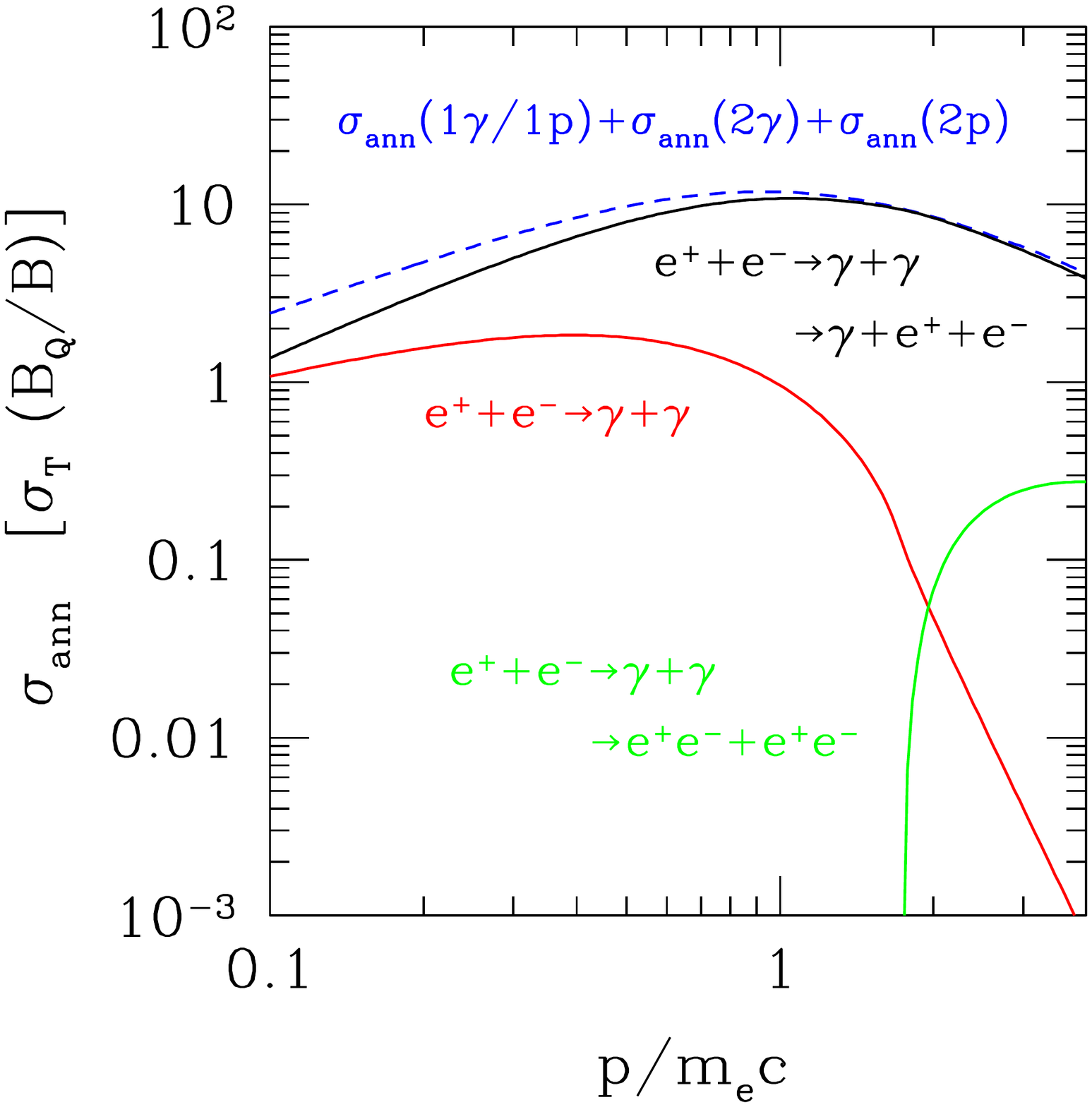}
\vskip -0.8in
\caption{Partial cross section for the annihilation of an electron and positron into a two-photon state
  in which (i) one photon immediately converts to a pair (black curve), (ii) both photons covert to pairs (green
  curve), (iii) neither photon is above the pair-conversion threshold (red curve).  Total cross section:
  blue dashed curve.  Like bremsstrahlung emission, channel (i) is logarithmically divergent in the minimum frequency
  of the emitted photon;  here we take $\hbar\omega_{\rm min} = 10^{-5}m_ec^2$.  The total cross section for
  annihilation bremsstrahlung is approximately twice the black curve, taking into account soft-photon emission
  by the final-state pair in the reaction $e^+ + e^- \rightarrow \gamma \rightarrow e^+ + e^-$ (Appendix \ref{s:soft}).}
\end{figure}\label{fig:sigma_ann}

\begin{figure}
  \epsscale{1.2}
  \vskip -0.3in
\plotone{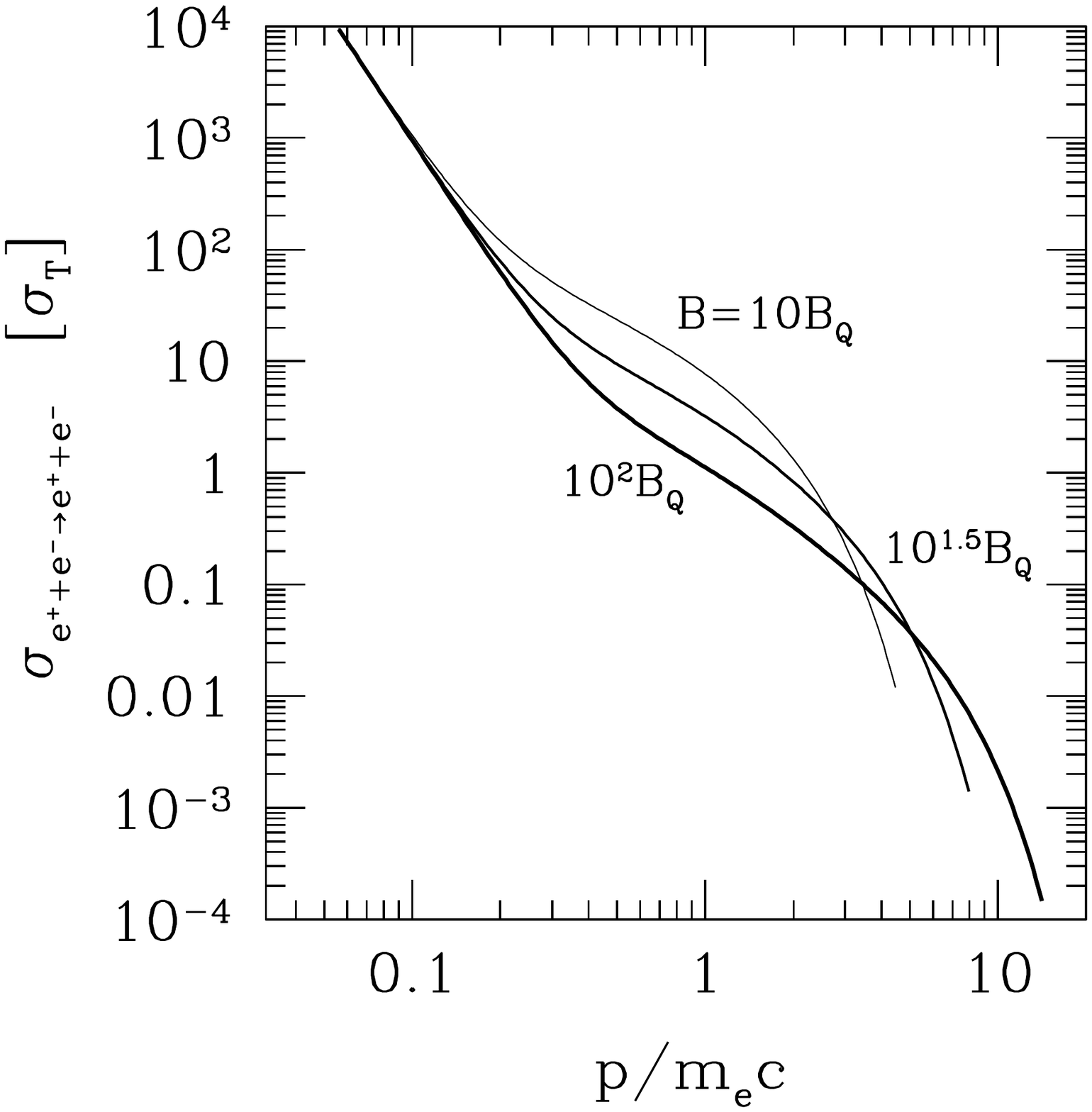}
\vskip -0.8in
\caption{Cross section for backscattering of $e^+$ and $e^-$ confined to the lowest
  Landau level, as a function of the longitudinal momenta $p_+ = -p_- = \pm p$, for various magnetic field
  strengths.  Curves are cut off at the momentum where excitation into the first excited Landau state becomes
  kinematically allowed.  Scattering is dominated by
  the Bhabha-like $t$-channel at $p \lesssim 0.24\,B_{15}^{1/3}\,m_ec$, where the cross section is independent of $B$;
  and by the resonant $s$-channel at higher momenta, where the cross section is ${1\over 2}$ that for single-photon
  pair annihilation.}
\end{figure}\label{fig:sigma_pp}

(iii) $e^+ + e^- \rightarrow \gamma + \gamma_{\rm soft} \rightarrow e^+ + e^- + \gamma_{\rm soft}$.
The two-photon annihilation spectrum shows an extended low-frequency tail very similar to
that of bremsstrahlung.  This feature will be central to our considerations.  In fact, the emission of the
soft photon may be viewed as a correction to the one-photon
annihilation of a pair, $e^+ + e^- \rightarrow \gamma$, which is kinematically allowed in the presence of the magnetic field.
Although one-photon pair annihilation is relatively fast, in a super-QED magnetic field the photon
will almost always reconvert to a pair before interacting with a second
particle.  One-photon annihilation therefore contributes mainly
to $e^+$-$e^-$ backscattering (\citealt{KT18}; Section \ref{s:current}).  In fact, when the electron and positron are
at least mildly relativistic, the backscattering cross section is dominated by the annihilation channel
($s$ channel), $e^+ + e^- \rightarrow \gamma \rightarrow e^+ + e^-$ (Figure \ref{fig:sigma_pp}).

We term this soft-photon emission process -- in which the pair is restored following the annihilation event --
annihilation bremsstrahlung.
The corresponding cross section has a low-frequency logarithmic divergence, like standard non-magnetic
bremsstrahlung, but with a much larger normalization.
The enhancement in the cross section, by a factor (\citealt{KT19}; Equation (\ref{eq:sigpm2}) below)
\be\label{eq:sigratio}
   {\sigma_{+-}(s)\over \sigma_{+-}(t)} \sim {4\pi \gamma^4\beta^3\over \alphem (B/\BQ)} = 76\gamma^4\beta^3\,B_{15}^{-1},
\ee
arises from the resonant nature of the $s$ channel in magnetic $e^+$-$e^-$ backscattering, in contrast with the Bhabha-like
momentum scaling of the $t$-channel contribution to the cross section.
(Here $\gamma$ and $\beta$ are the Lorentz factor and speed of the electron and positron
in the center-of-momentum frame.)
In other words, the emission of an additional soft photon involves an effective coupling
$\sim 4\pi (\BQ/B)$ as compared with $\alphem$ in ordinary bremsstrahlung emission.  The soft photon emissivity
is further enhanced by a factor 2 when the $s$ channel dominates $e^+$-$e^-$ backscattering, because a soft photon line
may also be attached to a final-state electron or positron (Appendix \ref{s:soft}).

The main conclusion here is that a pair gas with a narrow momentum distribution will emit an extended, power-law X-ray spectrum, similar to that
observed above $\sim 10$ keV in quiescent magnetars.  This arises from a single QED process, as just described, and without any need to invoke
a relativistic spread in $e^\pm$ energy, or multiple scatterings by warm pairs.  The $B^{-1}$ dependence in Equation (\ref{eq:sigratio})
plays an important role by weighting the annihilation bremsstrahlung emission away from the magnetar surface, at
a distance where the magnetic field has dropped to $\sim 5\,\BQ$.  In weaker magnetic fields, annihilations still take
place but the low-frequency spectral tail is substantially reduced (Section \ref{s:2gann}).

(iv) $\gamma + \gamma \rightarrow e^+ + e^-$.  The rate at which colliding photons convert into pairs is enhanced by a factor $B/\BQ$ as compared with an unmagnetized
vacuum, as originally derived by \cite{KM86}.  By contrast, the annihilation cross section is suppressed by the inverse of this factor,
due to the narrowing of the $e^\pm$ wavefunction in the directions transverse to an ultrastrong magnetic field
\citep{DB80}.  (The relation between these scalings is easily derived by examining detailed balance in
a thermal $e^\pm$ plasma.)  Collisions of soft photons with hard photons just below the single-photon conversion
threshold are kinematically allowed in a strong magnetic field, and the cross section is strongly enhanced compared with the case where the
photons have comparable energies.  This cross section can be written in a simple, compact form in the regime of magnetar-strength
magnetic fields \citep{KT18}.

(v) $e^+ + e^- \rightarrow e^+ + e^-$. In contrast with Bhabha scattering \citep{LLT4} there is no interference between the
$s$-channel and $t$-channel contributions
to the cross section for $e^+$-$e^-$ backscattering, meaning that the cross section may be written as a discrete sum
(\citealt{KT19}; Appendix \ref{s:soft})
\ba\label{eq:sigpm2}
\sigma_{+-} &\;=\;& \sigma_{+-}(t) + \sigma_{+-}(s) \nn
&\;=\;& {\pi r_e^2\over 4\beta^4\gamma^6} +
{\pi^2 r_e^2\over \beta\gamma^2\alphem(B/\BQ)}e^{-2\gamma^2\BQ/B}.\nn
\ea
Here $r_e = e^2/m_ec^2$ is the classical electron radius.  Equation (\ref{eq:sigpm2})
is plotted in Figure \ref{fig:sigma_pp}.   The absence of an interference term is directly related to the
existence of the 1-photon annihilation channel:  the intermediate-state photon is a real, not virtual, particle.
As described previously, the $s$-channel term in $\sigma_{+-}$ is strongly
enhanced for transrelativistic pairs compared with the Bhabha-like scattering ($t$-channel) term.  This enhancement is
easily explained by noting that $\sigma_{+-}(s)$ is essentially the cross section for {\it single-photon}
pair annihilation, differing only by a factor ${1\over 2}$ from the expression derived by
\cite{Wunner79} and \cite{DB80}.   (After the decay photon reconverts to a pair, the $e^\pm$ gain positive and negative
momenta with equal probabilities but only one sign corrresponds to backscattering.)
One sees in Figure \ref{fig:sigma_pp} that even transrelativistic $e^\pm$ backscatter strongly
through the $s$-channel when the magnetic field is $\sim 10-10^{1.5}\,\BQ$;  the $t$-channel dominates only
for $\beta \lesssim 0.24\,B_{15}^{1/3}$ in the center-of-momentum frame.
The relatively large value of this cross section
implies an enhanced collisionality and resistivity within a bulk pair plasma (Section \ref{s:current}).

\begin{figure*}
\epsscale{1.1}
\plottwo{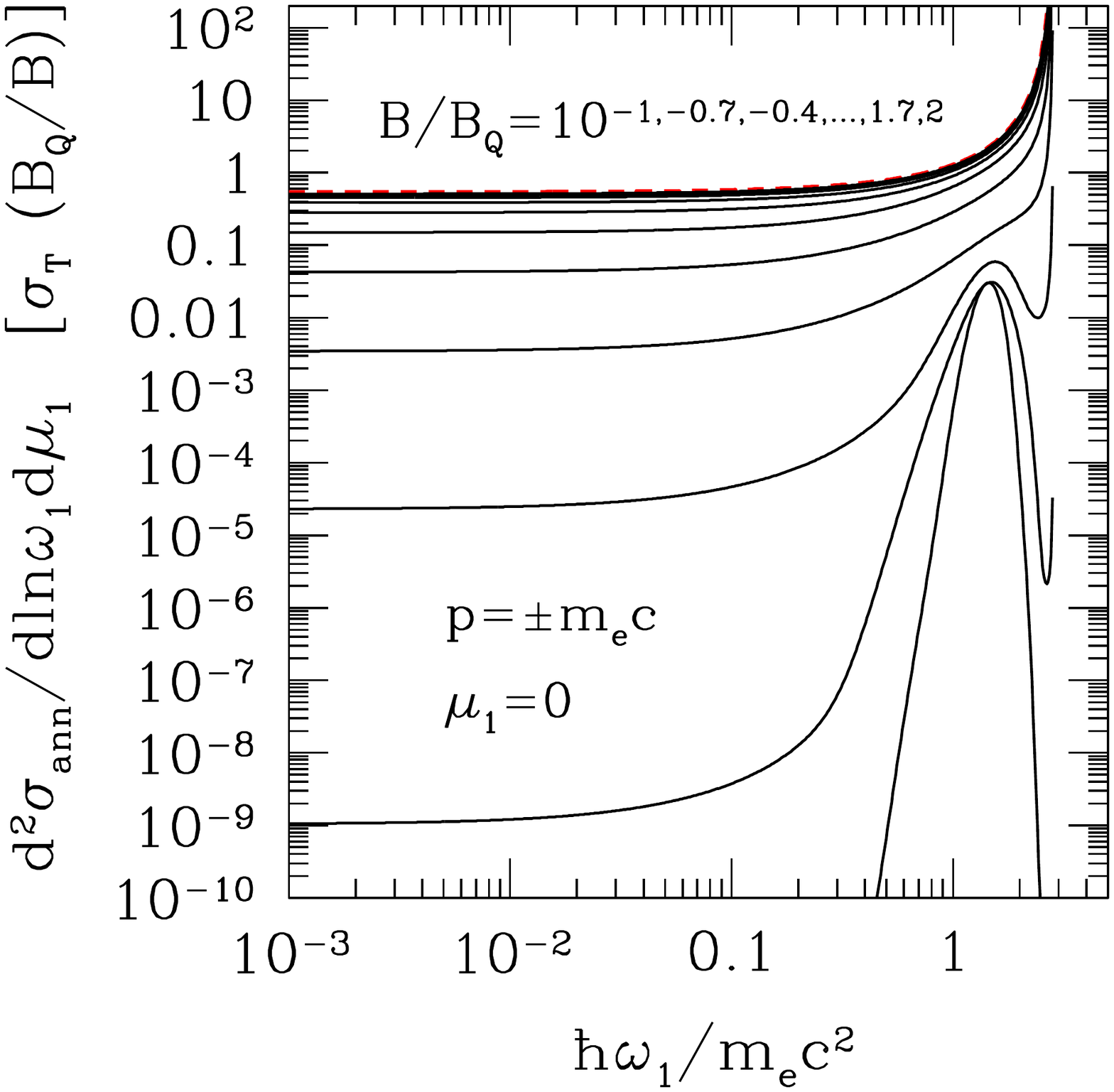}{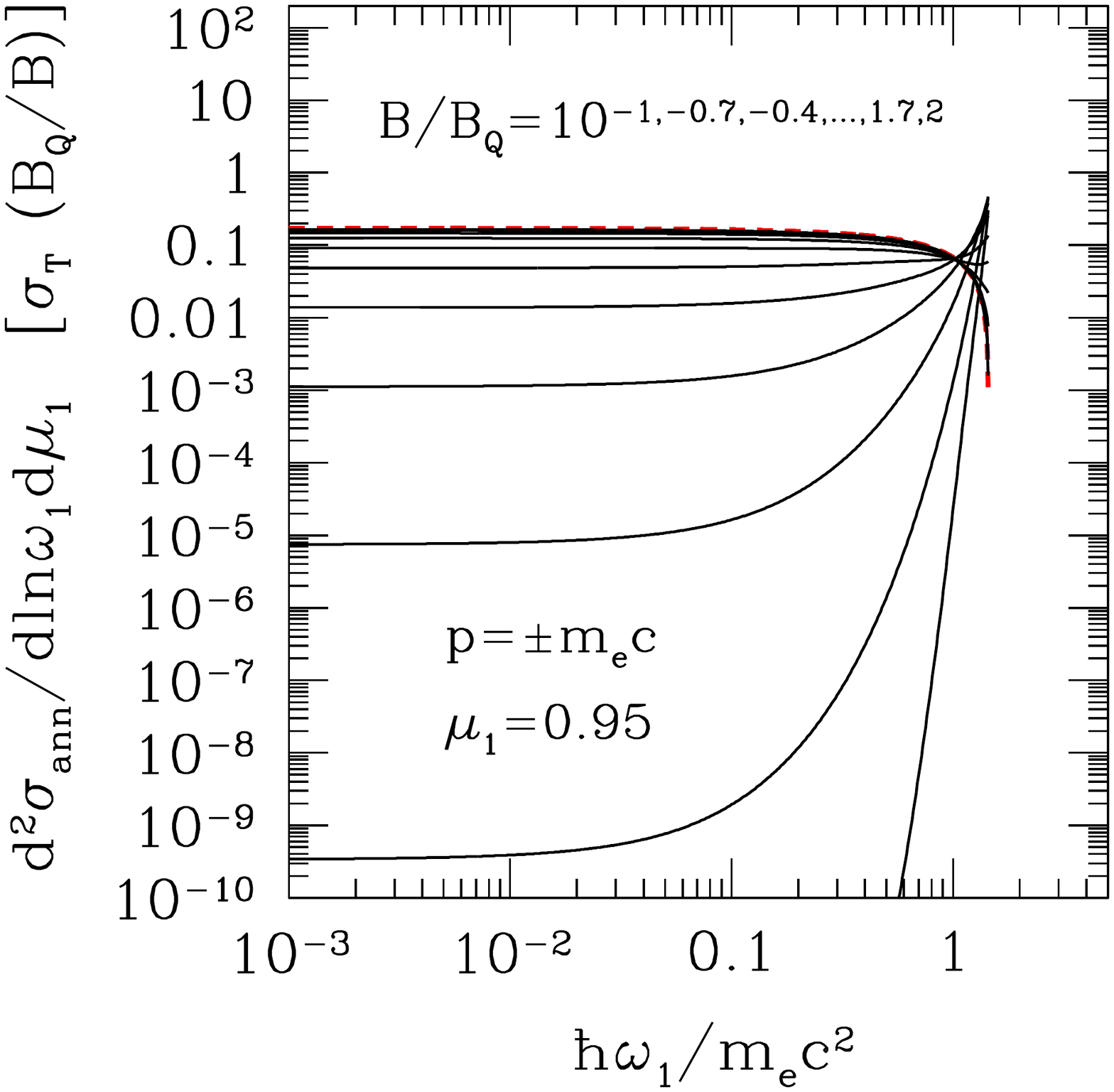}
\vskip -0.8in
\epsscale{0.57}
\plotone{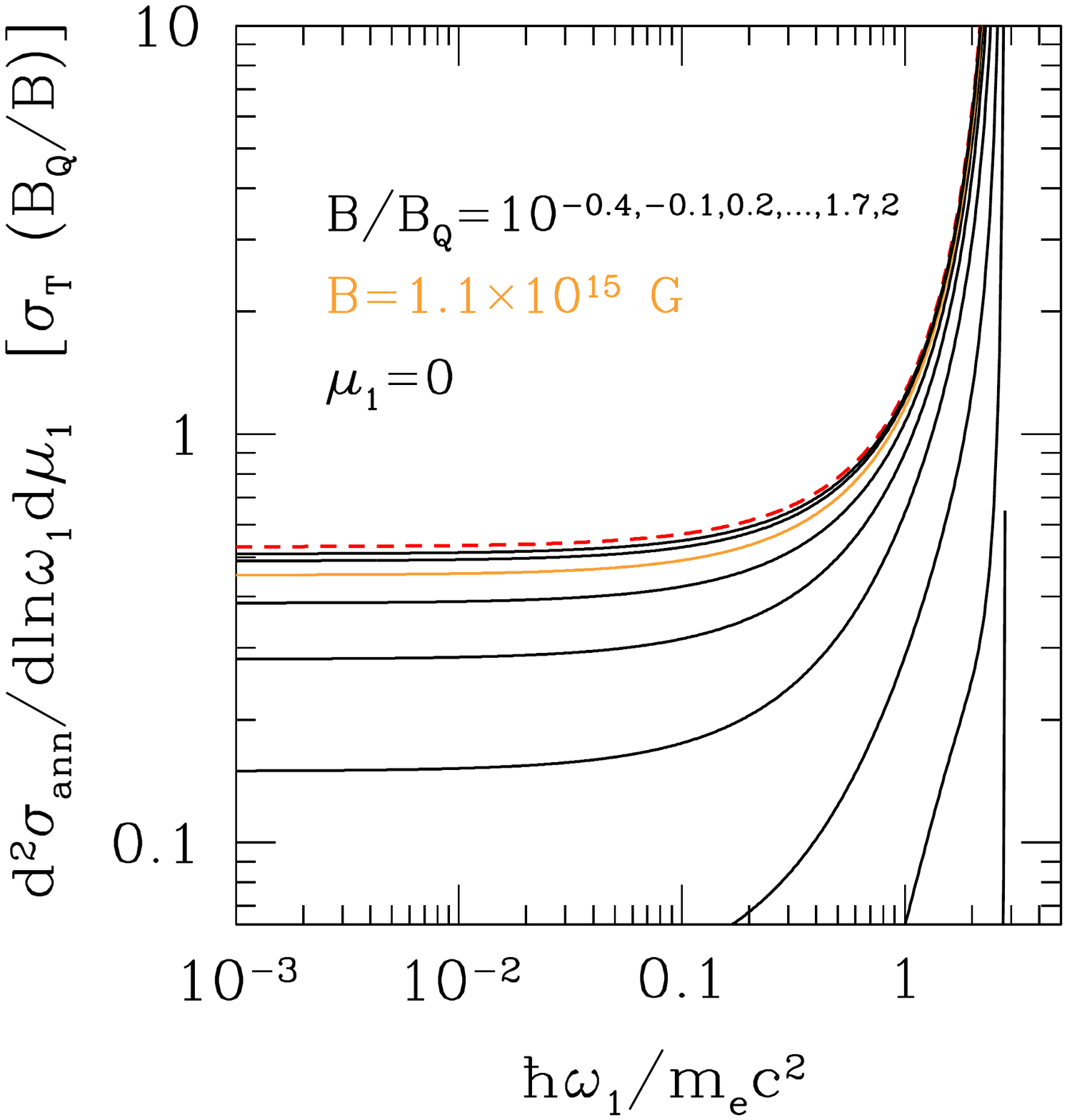}
\vskip -0.6in
\caption{Black curves:  differential cross section for production of a photon of frequency $\omega_1$ and direction cosine
  $\mu_1 = \cos\theta_1$ (measured with respect to ${\bf B}$) during the two-photon annihilation of an electron
  and positron from the lowest Landau level and with longitudinal momenta $p = \pm m_ec$.  Left panel:  $\mu_1 = 0$;  right panel:  $\mu_1 = 0.95$.
    Dashed red curve:  the strong-$B$ approximation of \cite{KT18}.  Black curves:  full result using matrix elements of \cite{DB80},
  for a range of magnetic fields extending from $10^{-1}\,\BQ = 4.4\times 10^{12}$ G (lowest)
  to  $10^2\,\BQ$ (highest), separated by $0.3$ in $\log_{10}B$.  
   The decay energy is shared roughly equally between the two photons when $B \lesssim 0.1\,\BQ$,
  corresponding to a narrow peak in the annihilation spectrum, as expected in a vacuum annihilation.  A low-frequency tail develops rapidly
  as $B$ rises above $\BQ$.  Third panel: expanded view demonstrating that the strong-$B$ approximation
  is accurate to better than a factor 2 when $B \gtrsim 5\,\BQ$.}
\end{figure*}\label{fig:annihil}

\subsection{Two-photon Annihilation Spectrum}\label{s:2gann}

Figure \ref{fig:annihil} shows how the soft, flat component of the annihilation spectrum emerges as $B$ rises above
$\BQ$.  The two-photon spectrum is concentrated in a line-like feature at $\hbar\omega \sim m_ec^2$ in weaker magnetic fields,
and does not differ substantially from the vacuum decay spectrum.  When $B\gtrsim 5\,\BQ \sim 2\times 10^{14}$ G there
is excellent agreement of the full cross section and the simple result of \cite{KT18}, given in Equation (\ref{eq:sigma_ann_m1m2})
below, which was derived in the regime $10\alphem^{-1}\BQ \gg B \gg \BQ$.
The full cross section is computed by performing an extended sum over the intermediate-state $e^\pm$ Landau level using
the matrix elements of \cite{DB80}.

The distribution of photons emitted in the center-of-momentum frame
with direction cosines $\mu_1 = \cos\theta_1 = \hat k_1\cdot\hat B$, $\mu_2 = -\mu_1\omega_1/\omega_2$ and frequencies $\hbar\omega_1 = 2\gamma
m_ec^2 \mu_2 / (\mu_2 -\mu_1)$, $\hbar\omega_2 = 2\gamma m_ec^2 - \hbar\omega_1$, is obtained from 
\be\label{eq:sigma_ann_m1m2}
   {d^2\sigma_{\rm ann}\over d\mu_1 d\mu_2} = {2\pi r_e^2\over B/\BQ} {\beta\over\gamma^2}
   {|\mu_1 - \mu_2| (1-\mu_1^2)(1-\mu_2^2)\over |\mu_1 \mu_2| [(1-\mu_1\mu_2)^2 - \beta^2(\mu_1-\mu_2)^2]^2}.
   \ee
Following the treatment in \cite{KT18}, we set to unity the angle-dependent vertex factors
$e^{-(\hbar\omega/m_ec^2)^2(1-\mu^2)\BQ/2B}$ associated with emission or absorption of a photon of
frequency $\omega$ and direction cosine $\mu$.  (Even in the case of a photon near the threshold for pair conversion, these do not
introduce a significant correction in a super-QED magnetic field, and their omission greatly facilitates
the construction of probability arrays in the MC code.)

\begin{figure}
  \epsscale{1.2}
  \vskip -0.3in
\plotone{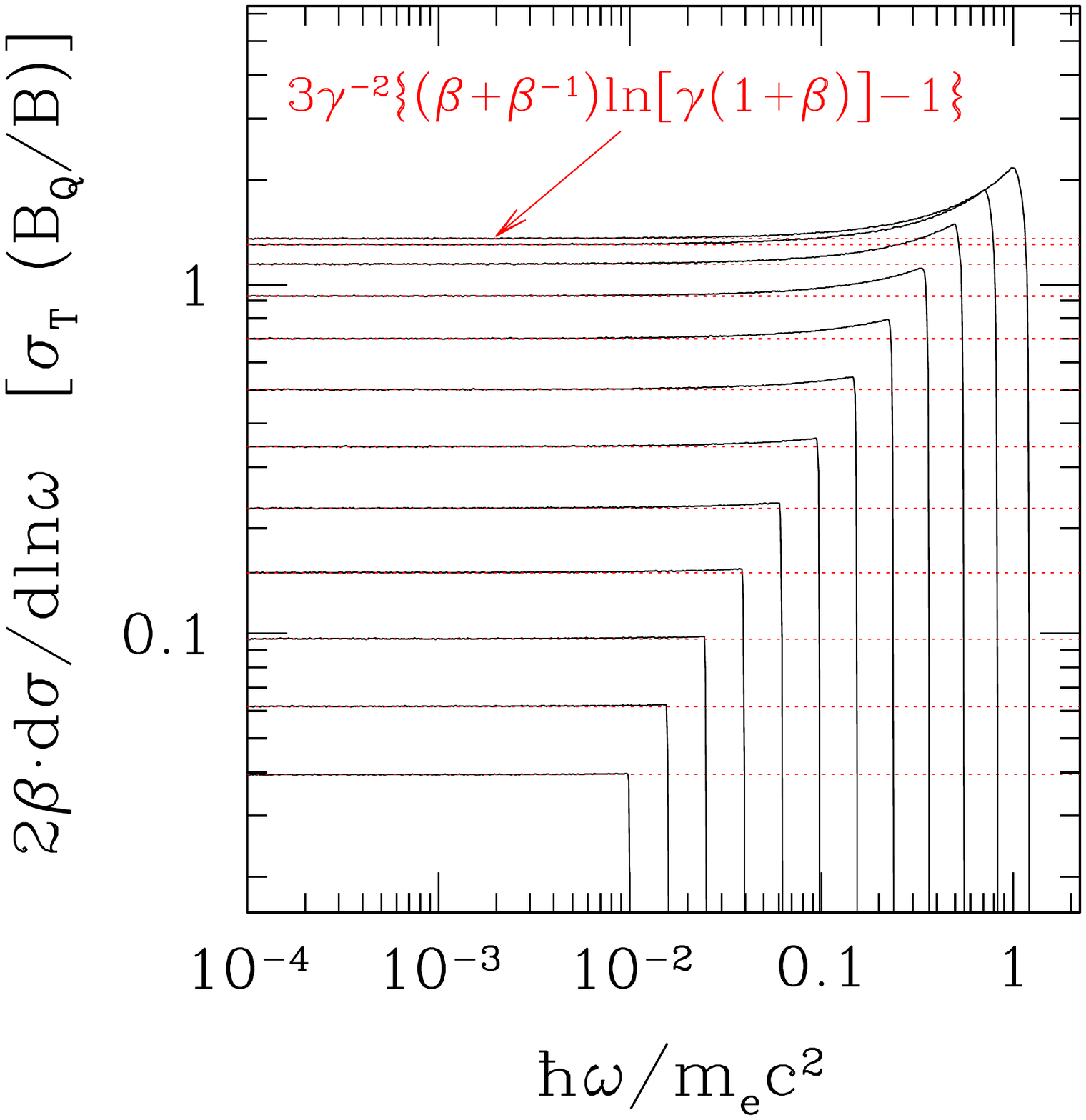}
\vskip -0.9in
\plotone{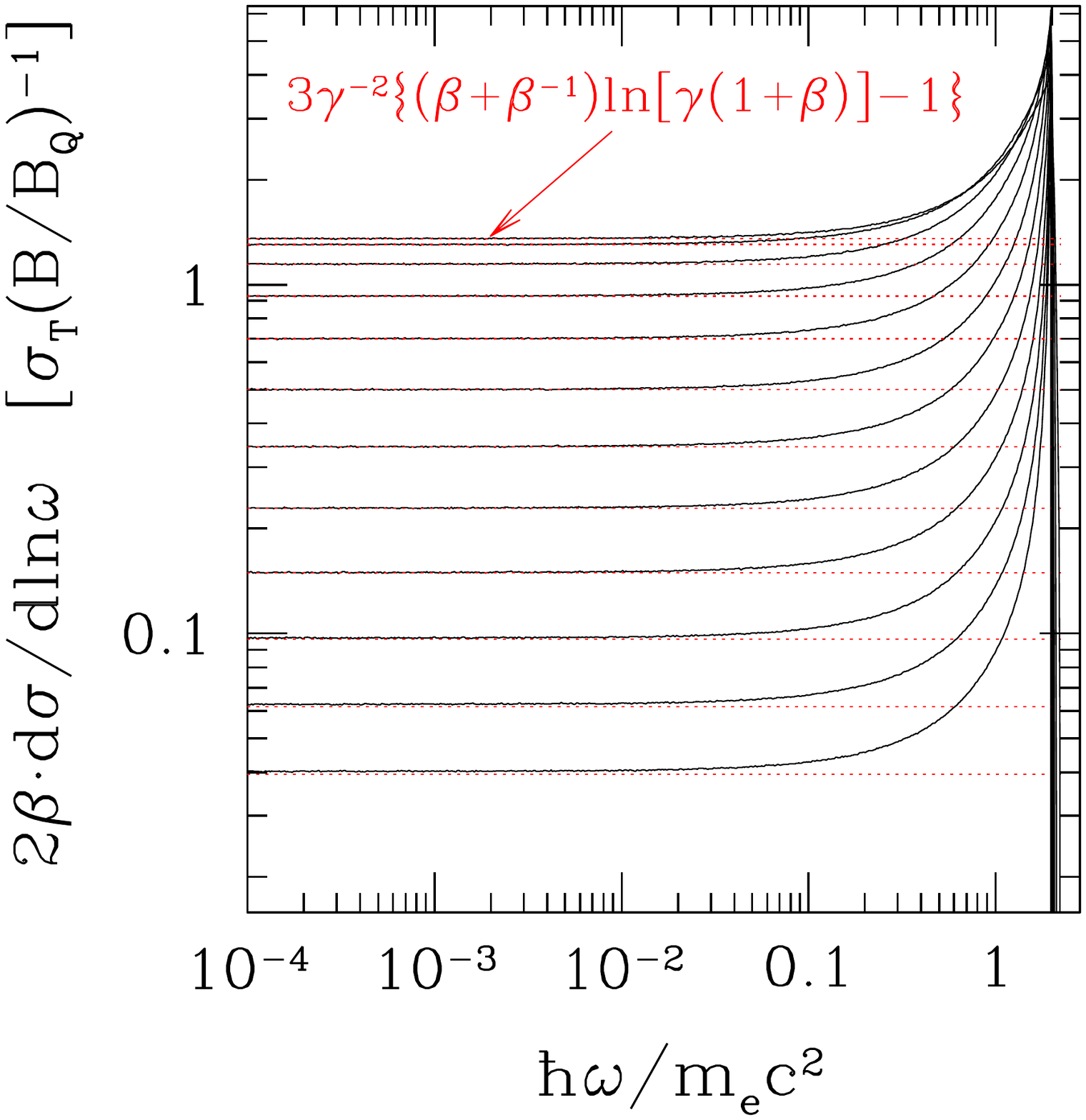}
\vskip -0.6in
\caption{Frequency distribution of photons produced by the annihilation of an electron and positron with
  longitudinal momenta $\pm p$, with $p/m_ec = 10^{-1,-0.9,-0.8,...0,0.1}$ (bottom to top black curves).
  Red dotted curves mark the
  low-frequency analytic approximation (\ref{eq:lowfreq}) to the differential cross section.
  Top panel:  result
  restricted to the channel with one final-state photon below pair-creation threshold.  Bottom panel: the
  full distribution with one or both of the final-state photons below threshold.}
\end{figure}\label{fig:ann_spec}

Converting one direction cosine to the frequency of an outgoing photon and integrating over the second
direction, we find in the soft-photon regime $(\omega_1 \ll \omega_2)$
\ba\label{eq:lowfreq}
2\beta\cdot \omega_1 {d\sigma_{\rm ann}\over d\omega_1} &=& {8\pi r_e^2\over \gamma^2\,B/\BQ}
\left\{\left(\beta + \beta^{-1}\right)\ln[\gamma(1+\beta)] - 1\right\}\nn
&\equiv& {8\pi r_e^2\over B/\BQ}f(\beta).
\ea
The full emission spectrum resulting from two-photon annihilation is shown in Figure \ref{fig:ann_spec},
along with the low-frequency approximation (\ref{eq:lowfreq}).  The left panel shows the 1-$\gamma$/1-p channel
in isolation, which is seen to dominate the low-frequency component, but cuts off at an energy
$\hbar\omega_1 = 2(\gamma-1) m_ec^2$.  For a given emission direction $\mu_1$, the boundary between
the 1-$\gamma$/1-p and 2-$\gamma$ final states lies at \citep{KT18}
\be
{|\mu_2|\over \mu_1} < {\gamma\sqrt{1+(\gamma\beta\mu_1)^2} - 1\over 1+\gamma^2\mu_1^2} \quad\quad ({\rm one~pair~produced});
\ee
whereas the threshold of the two-pair final state is (when $\gamma > 2$)
\be
{|\mu_2|\over\mu_1} > {1\over \gamma\sqrt{1-\mu_1^2} - 1};  \quad \mu_1 < \sqrt{1 - {4\over\gamma^2}}.
\ee

As described above and in Appendix \ref{s:soft}, the emission of a soft photon in combination with a hard, pair-converting photon can
be viewed as a soft-photon correction to $e^+$-$e^-$ backscattering in the $s$ channel.  The total emission spectrum
produced by $e^+$-$e^-$ collisions
includes soft photons that are emitted by the final-state pair. 
In the MC code, this contribution is approximated as the
1-$\gamma$/1-p emission spectrum in the left panel of Figure \ref{fig:ann_spec}, which is exact in the soft-photon limit.
In other words, the total emission spectrum is taken to be the sum of the right and left panels of Figure \ref{fig:ann_spec}:
\ba\label{eq:sigma_tot}
         {d^2\sigma_{\rm em}\over d\mu_1 d\omega_1} &=& {d^2\sigma_{\rm ann}\over d\mu_1 d\omega_1}\biggr|_{\omega_1 < \omega_{\rm th}(\mu_1)} \nn
         &+& {d^2\sigma_{\rm ann}\over d\mu_1 d\omega_1}\biggr|_{\omega_1 < \omega_{\rm th}(\mu_1),~\omega_2 > \omega_{\rm th}(\mu_1)}.
\ea
(Here $\omega_{\rm th} = 2m_ec^2/\sin\theta$.)         
The first term is identical to the second when one photon is emitted above the threshold for pair conversion.
In other words, the hard X-ray output of a trans-relativistic pair
gas is twice that expected from the two-photon decay cross section (\ref{eq:sigma_ann_m1m2}).

The total cross section for soft-photon emission is, to logarithmic accuracy,
\be\label{eq:sigsoft}
2\sigma_{\rm ann}(1\gamma/1{\rm p}) = {16\pi r_e^2\over B/\BQ} {f(\beta)\over 2\beta} \ln\left({\omega_{\rm max}\over\omega_{\rm min}}\right).
\ee
Here, $\omega_{\rm min}$ and $\omega_{\rm max} \sim 2(\gamma-1)m_ec^2/\hbar$ are the minimum and maximum emitted frequencies in the 1-$\gamma$/1-p channel.
The function $f(\beta)$, defined in Equation (\ref{eq:lowfreq}), has the low-velocity asymptote $f(\beta) \rightarrow 4\beta^2/3$.  

In this paper, we neglect the emission of soft photons associated with $e^+$-$e^-$ backscattering in the $t$ channel --
the process most closely analogous to standard bremsstrahlung -- since we
are focused on pair plasmas warm enough for the $e^+$-$e^-$ cross section to be dominated by the annihilation channel.
(The soft-photon spectrum associated with each channel is proportional to the scattering cross section -- see
Equation (\ref{eq:soft}) in Appendix \ref{s:soft} and the surrounding discussion.)

\section{Monte Carlo Method}\label{s:monte}

We have developed a MC code that describes the emission, scattering, and re-absorption of 
X-rays and gamma-rays by a transrelativistic $e^\pm$ cloud.  Emission is by two-photon annihilation, including both the
1-$\gamma$/1-p and 2-$\gamma$ final states, using the prescription of Equation (\ref{eq:sigma_tot}).
Soft-photon emission through the $t$-channel (which is closely analogous to ordinary bremsstrahlung)
is subdominant for the $e^\pm$ energies considered ($\beta \gtrsim 0.1$), and is ignored.

The plasma is embedded in an
ultra-strong magnetic field and has a cartesian, slab geometry, with the magnetic field
running parallel to the slab.  It is described by three parameters:  (i) the magnetic flux density
$B$, (ii) the total column density $N_\pm = N_+ + N_- = 2N_-$ across the slab (normalized by the dimensionless parameter
$N_\pm \pi r_e^2$), and (iii) the thermal
velocity of the pairs, which is restricted to a single cartesian direction (parallel to ${\bf B}$)
and for ease of computation is taken to be $\pm \beta c \hat B$ for both electrons
and positrons.  The drift speed that supports the current is expected to be a small fraction of $\beta$
(Section \ref{s:current}), and so any directional imbalance is ignored.  

Scattering of photons off $e^\pm$ is described by the strong-$B$ cross
section derived in \cite{KT18} (Equations (27) and (28)).  We allow both for scattering into non-pair converting
and pair-converting final states ($\hbar\omega <$/$> 2m_ec^2/\sin\theta$), with the latter outcome implying immediate
absorption of the photon.  Finally,
two-photon pair creation is handled in a two-step process, by (i) summing the occupancies of successive propagating and scattering
photons in the $\omega$-$\mu$ space, and (ii) constructing from this two-parameter density distribution 
an absorption coefficient for each $\omega$-$\mu$ bin using the $\gamma$-$\gamma$ collision cross section
of \cite{KT18}. 

We now describe each step in the MC procedure in more detail.  Direction cosine $\mu_1$ of the emitted
photon is divided typically into $N_\mu = 2^5$ bins uniformly spaced between $-1$ and $1$, with $\mu_1$ evaluated at
the bin center.  In contrast with non-magnetic MC scattering codes, it is necessary to adopt a $\mu$-dependent
frequency binning, since the scattering cross section peaks strongly just below the single-photon pair conversion
threshold, which depends on propagation angle (Figure \ref{fig:tauscatt}).  We choose (for $1 \leq i < N_\omega \sim 2^{10}$
and $1 \leq j \leq N_\mu$)
\be
{\omega_{1,i}(\mu_{1,j})\over\omega_{\rm min}} =  { e^{(i-1)\delta}\over 1 + [\omega_{\rm min}/\omega_{\rm 1, max}(\mu_{1,j})]
  [e^{(i-1)\delta} - 1]},
\ee
where $\omega_{\rm min}$ is a low cutoff frequency and
\be\label{eq:ommax}
\hbar\omega_{1,\rm max}(\mu_1) = {\rm min}\left[{2\gamma m_ec^2 \over (1+\mu_1)}, \;{2m_ec^2\over (1-\mu_1^2)^{1/2}}\right].
\ee
Here, $\omega_{1,i}(\mu_{1,j})$ is the lower boundary of the $i^{\rm th}$ frequency bin in the $j^{\rm th}$ $\mu_1$ bin.
The left term in Equation (\ref{eq:ommax}) is the energy of photon 1 evaluated at $|\mu_2| = 1$.

{\sl Emission.}  As a first step, a cumulative probability distribution is constructed for the frequency
of the emitted photon in each $\mu_1$ bin, which can be done by a (somewhat cumbersome) analytic integration of the differential cross
section defined in Equation (\ref{eq:sigma_tot}),
\be
P_2(\mu_1,\omega_1) = {1\over d\sigma_{\rm em}/d\mu_1(\mu_1)}\int_{\omega_{\rm min}}^{\omega_1} d\omega_1' {d^2\sigma_{\rm em}\over d\mu_1d\omega_1'},
\ee
where
\be\label{eq:sigmaem}
   {d\sigma_{\rm em}\over d\mu_1}(\mu_1) \equiv
   {dP_1\over d\mu_1}(\mu_1)\cdot \sigma_{\rm em} = \int_{\omega_{\rm min}}^{\omega_{1,\rm max}(\mu_1)} d\omega_1' {d^2\sigma_{\rm em}\over d\mu_1d\omega_1'}.
\ee
Here $\sigma_{\rm em}$ is the sum of the total cross section for the combined 1-$\gamma$/1-p and 2-$\gamma$ channels, and
the cross section for the 1-$\gamma$/1-p channel alone (see Equation (\ref{eq:sigma_tot})).
The cumulative $\mu_1$ probability distribution $P_1(\mu_{1,j})$ must be evaluated by a numerical integration of $P_2$ over $\mu_1$ bins.

A photon is initiated first by drawing a random number and using the function $P_1$ to determine from this the 
bin-centered value of $\mu_1$.  Then a second random number is drawn to determine the bin-centered frequency, using the
function $P_2$.

{\it Scattering.}  At low photon frequencies, scattering off $e^\pm$
is well approximated by the classical cross section $d\sigma_{\rm es}/d\mu_2 = 2\pi r_e^2
(1-\mu_1^2)(1-\mu_2^2)$, where now labels 1 and 2 represent the incident and scattered photons.   However, the cross
section rises dramatically as $\omega_1$ approaches the single-photon pair conversion threshold, due to a $u$-channel
resonance between the incident electron, a virtual positron, and the final-state photon.  Figure \ref{fig:tauscatt} shows the
angle-averaged cross section for a photon as a function of incident frequency.   The concentration of the two-photon
annihilation energy in a single photon gives this resonance an important effect on radiative energy escape from
the pair plasma.

Scattering is handled in an analogous manner to emission, in two steps.  (i) A probability distribution $P^{\rm scatt}_1(i,j,\mu_2)$
is constructed for the final direction cosine $\mu_2$ in each $\omega_1$-$\mu_1$ bin of the incident photon (labelled $(i,j)$).
This distribution describes scattering in the rest frame of the $e^\pm$, and is obtained from the cross section
presented in \citealt{KT18}.   (ii) The total scattering cross section is obtained by numerical integration
in each $\omega_1$-$\mu_1$ bin.  Since the momentum distribution of the $e^\pm$ is assumed to be peaked at $\pm p \equiv
\pm \gamma\beta m_ec$, this cross section must be evaluated only for two pairs of direction cosine and
frequency $(\mu_{1,r},\omega_{1,r})^\pm$, each Lorentz boosted to the charge rest frame.
A scattering event is determined first by evaluating the two scattering depths along the photon ray to the edge of the scattering box,
\be
\tau_{\rm es}^\pm(\mu_1,\omega_1) = {n_\pm\over 2}{\Delta l_\perp\over |\mu_1|}
(1\mp\beta\mu_1)\sigma_{\rm es}(\mu_{1,r}^\pm,\omega_{1,r}^\pm).
\ee
Here $n_\pm = n_+ + n_-$ is the total space density of electrons and positrons, and $\Delta l_\perp$ the coordinate displacement
  transverse to ${\bf B}$.
This procedure is repeated separately for final states in which the photon is, or is not, above the threshold for pair conversion.
Drawing a single random number allows us to determine, first, which if any of these four scattering channels is activated and then,
if one is, the length along the ray at which the scattering takes place.  A second random number then determines the bin-centered
direction cosine of the outgoing photon in the initial $e^\pm$ rest frame.  A final Lorentz boost gives the outgoing direction
cosine and frequency in the `lab' frame.

The photon frequency is determined by exact energy conservation during these successive boosts, with only the direction cosine
discrete.  As a result, it is possible for a photon that is below the pair-conversion threshold in one frame to exceed it in another
frame.  The fineness of the chosen frequency binning near the conversion threshold allows such events to remain rare
(less than a percent of scattering-induced pair conversion events are such spurious events).

The frequency and direction cosine are recorded when a photon escapes the scattering box, and its energy is counted as
a loss to the pair gas.  By contrast, its energy is assumed to be returned to the gas following an internal conversion event.

{\it Photon-photon collisions.}  The most challenging process to model is two-photon pair conversion, $\gamma + \gamma \rightarrow
e^+ + e^-$.  One requires a binned density field for the target photons, which can be built up during successive MC steps.
Furthermore, the magnetized cross section is very large when one of the colliding photons has a low energy
\citep{KM86, KT19}, and we follow photons over a wide range ($\sim 10^4$) in frequency.
We first inject a small fraction ($\sim 2^{-10}$) of the total number ($\sim 2^{30}$) of photons and follow them without
this opacity source.  The resulting photon density field is used to evaluate the next batch of photons, now with
the $\gamma$-$\gamma$ opacity strongly suppressed.   The suppression factor is gradually relaxed over $\sim 20$ steps
of the same size, yielding a calibrated opacity that is used for the remaining trials.

The normalization of the photon density field proceeds as follows.  A density $n_\pm$ of pairs results in a volumetric
emission rate\footnote{The factor of 2 multiplying $\sigma_{\rm ann}(1\gamma/{\rm 1p})$ counts soft photons emitted by final-state $e^\pm$
in the annihilation channel of $e^\pm$ backscattering (Appendix \ref{s:soft}); the factor 2 multiplying $\sigma_{\rm ann}(2\gamma)$
counts the emission of two photons below the pair-conversion threshold.}
\be
\dot n_\gamma = 
2\beta c \left[2\sigma_{\rm ann}(1\gamma/{\rm 1p}) + 2\sigma_{\rm ann}(2\gamma)\right]\,{n_\pm^2\over 8}.
\ee
The last factor in this expression counts collisions of left-moving $e^+$ with right-moving $e^-$, and vice versa.
A photon traversing a ray of length $\Delta l$ has an implied density
\be
   {\Delta n_\gamma\over n_\pm} =  {\Delta l \over L} \beta \tau_{\rm T}\,
   {\sigma_{\rm ann}(1\gamma/{\rm 1p}) + \sigma_{\rm ann}(2\gamma)\over 2\sigma_{\rm T}},
\ee
where $\tau_{\rm T} = \sigma_{\rm T} N_\pm = \sigma_{\rm T} n_\pm L$ is the Thomson depth across the plasma slab (of width $L$).
Normalizing the photon density to
the pair density in this way allows us to construct a $\gamma$-$\gamma$ optical depth that is proportional to the scattering
depth.   The strong-field limit for the photon collision cross section $\sigma_{\gamma\gamma}$ is used (see Equation (7) in \citealt{KT19}, Erratum).
Constructing the opacity involves a sum over the (bin-centered) direction cosine and frequency of the target photon,
\be
\kappa_{\gamma\gamma}(\mu_1,\omega_1) = \sum_{i,j} \sigma_{\gamma\gamma}(\mu_1,\omega_1,\mu_{2,j},\omega_{2,i})
      {\Delta n_\gamma(i,j)\over N_{\rm ann}}.
\ee
Here, the array $\Delta n_\gamma$ is summed over all injected photons and then divided by the implied number of
annihilations,
$N_{\rm ann} = N_\gamma [2\sigma_{\rm ann}(1\gamma/{\rm 1p}) +
  \sigma_{\rm ann}(2\gamma) + \sigma_{\rm ann}({\rm 2p})] /
[2\sigma_{\rm ann}(1\gamma/{\rm 1p}) + 2\sigma_{\rm ann}(2\gamma)]$.

The decision whether a given photon will scatter of an $e^\pm$ (moving with either positive or negative momentum)
into a state that is either above or below threshold for pair conversion, or instead will collide with another photon,
involves a choice between 5 outcomes.  We do not distinguish between photon collisions involving different summed
photon energy, since in all cases the photon energy is returned to the pair gas.  Each of these outcomes
has a probability $\kappa_i$ per unit length, $1\leq i \leq 5$.
The drawing of a single random number ${\cal R}$ allows us to decide, in two steps, which if any of the 5 processes
is activated.  The first step involves finding the process $i$ for
which ${\cal R} \subset (P_{i-1}, P_i)$, where $P_i = \sum_{j=1}^i \kappa_j / \sum_{j=1}^5 \kappa_j$
and $P_0 = 0$.  In the second step, we compute the scattering/absorption probability $1 - e^{-\kappa_i l}$ for process $i$ in isolation,
where $l$ is the ray path length to the plasma edge.   If $1 - e^{-\kappa_i l} > ({\cal R} - P_{i-1})/(P_i-P_{i-1})$,
then the scattering/absorption event occurs; otherwise, the photon leaves the box and its frequency and direction cosine
are recorded.  If a pair conversion of the photon is triggered, then its propagation is terminated;  otherwise, the preceding
step is repeated, using the updated energy, direction cosine, and starting position (scattering coordinate) of the photon.

Finally, we noted that a certain number of two-pair annihilation events is implied for each batch of injected photons.  These
events are not followed in the MC but they are compared as a net source of pairs with the scattering-assisted
and two-photon channels (Section \ref{s:results}).

\begin{figure}
\epsscale{1}
\plotone{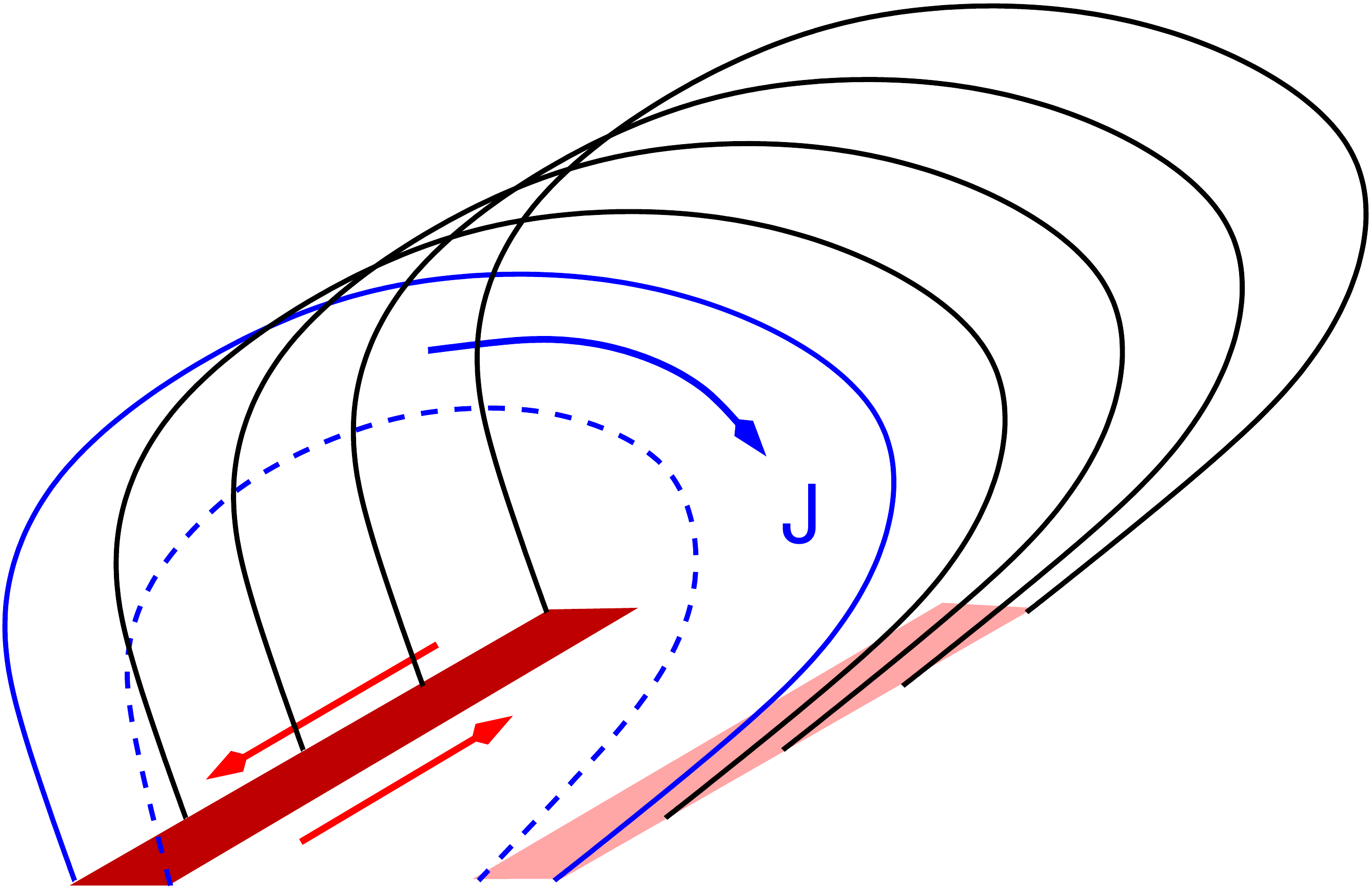}
\caption{Sheared magnetic arcade loaded with transrelativistic $e^\pm$ and anchored at both ends within the magnetar crust.
  Localized plastic flow along a fault-like feature drives the magnetospheric current, as seen in the global
  simulations of \cite{TYO17}.}
\end{figure}\label{fig:fluxtube}

\section{Current Flow and Ohmic Heating}\label{s:current}

The $e^\pm$ plasma considered here is cool enough to remain confined to the lowest Landau state in a super-QED magnetic field.
The MC results presented in Section \ref{s:results} reveal that the competition between internal pair creation and annihilation
pushes the thermal momentum to a transrelativistic value $p \sim \pm (0.5-1)m_ec$.  The magnetic field is assumed
to be sheared or twisted (Figure \ref{fig:fluxtube}).   The current system
is nearly force-free, given the low plasma kinetic pressure, and supports a current density ${\bf J}$ flowing nearly parallel to ${\bf B}$.

We will demonstrate (Section \ref{s:eqdepth}) that the baseline collisional heating of the current-carrying plasma (driven by $e^+$-$e^-$ backscattering)
can be compensated by radiative cooling, but only if the $e^\pm$ density
is 10-20 times the minimum $n_{\pm,\rm min} = |J|/\beta ec$ that will support the current.   Then the differential
$e^+$-$e^-$ drift speed is much smaller than $\beta c = p/\gamma m_e$, and a symmetric momentum distribution is a reasonable first
approximation for modelling the radiative emissions.  

\begin{figure}
  \epsscale{1.2}
  \vskip -0.3in
\plotone{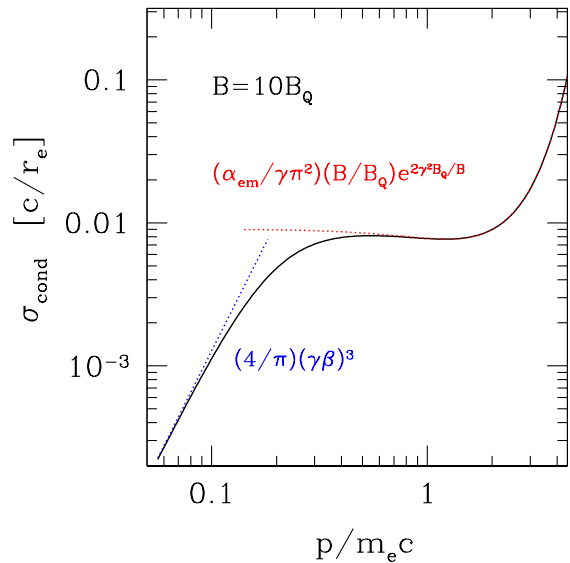}
\vskip -0.8in
\caption{Drude electrical conductivity of a pair plasma embedded in a magnetic field $B = 10\,\BQ$, as a function
  of the thermal momentum.  (Distribution function peaked at $\pm p$, as assumed in the MC code.)}
\end{figure}\label{fig:cond}

\begin{figure}[t]
  \vskip -0.3in
\epsscale{1.2}
\plotone{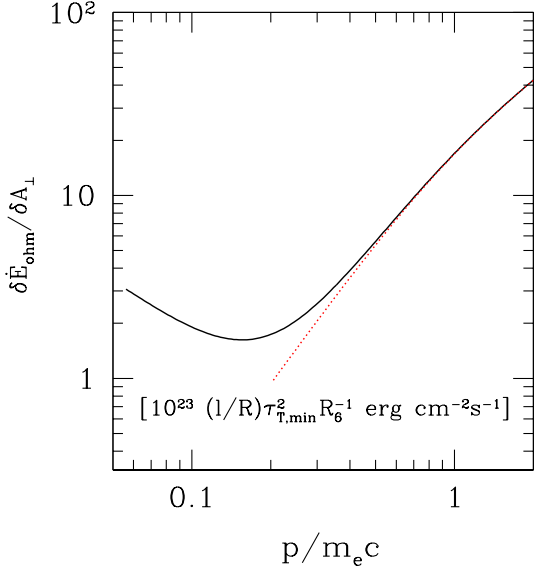}
\vskip -0.8in
\caption{Ohmic heating rate integrated over the volume of a closed magnetic flux bundle, per unit area of
  a single footpoint, as a function of the thermal momentum of the embedded $e^\pm$.  Surface magnetic
  field $B(R) = 10\,\BQ$.  The ohmic heating is nearly uniform per unit length of the loop,
  except for $p$ approaching the upper range plotted.  Red dotted curve:  approximation given in Equation (\ref{eq:edotohm}).
  Following the analytic treatment of ohmic dissipation in the text, we have approximated the exponential
  factor in $\sigma_{+-}(s)$ as unity.}
\end{figure}\label{fig:ohm}

The strength of the magnetic shear is characterized by a scattering depth $\tau_{\rm T,min}$ associated with the minimum pair density
$n_{\pm,\rm min}$.  Writing  $J = c B/4\pi l_{\rm shear}$ and taking into account that $J/B \sim$ constant in a
force-free magnetic field, the shear scale $l_{\rm shear} = B/|\bnabla\times {\bf B}|$ is also a constant along
a sheared magnetic flux bundle.  The scattering depth is normalized as
\be
\tau_{\rm T,min} = {R\over 2}\sigma_{\rm T} n_{\pm,\rm min}(R) \sim {R\over 3l_{\rm shear}} {\alphem\over\beta} {B(R)\over \BQ},
\ee
where $R$ is the stellar radius.
As defined, $\tau_{\rm T,min}$ coincides with the magnetic field-aligned integral from the magnetar surface to the top
of the flux bundle, $\int_0^{l(R_{\rm max})} \sigma_{\rm T} n_{\pm,\rm min}(l) dl$,
when the field is nearly radial at the magnetar surface and extends to a radius $R_{\rm max} \gtrsim 3R/2$.
We will compare this with the depth through the magnetospheric plasma, normalized similarly as
\be\label{eq:tauTR}
\tau_{\rm T}(R) \simeq {1\over 2}\sigma_{\rm T} n_\pm(R)R.
\ee

The quantity $\tau_{\rm T,min}$
can approach or exceed unity when the surface magnetic field $B(R) \sim 10^{15}$ G  and the shear scale is small ($< R/2$).
In fact, the shear pattern in an active plastic deformation zone (Figure \ref{fig:fluxtube}) may
not be smooth, being composed of modes with a small amplitude $\delta B \ll B$ but a high wavenumber $k_\perp \gg ({\rm km})^{-1}$,
producing a high absolute current density $B(R)/l_{\rm shear} \rightarrow |k_\perp \delta B(R)|$.   The plasma equilibrium
described here depends on the magnitude of the current, not its sign.

The Drude conductivity of the $e^\pm$ plasma is (Figure \ref{fig:cond})
\be\label{eq:sigmacond}
\sigma_{\rm cond} = {e^2\over \beta\gamma^3 m_ec \sigma_{+-}}.
\ee
Here, as in the MC code, we adopt a characteristic thermal momentum; the inverse factor of $\gamma^3$ represents
the longitudinal relativistic mass.   The normalization of $\sigma_{\rm cond}$ takes into account
that $e^+$ can only backscatter off $e^-$ in a one-dimensional plasma.  Mutual scatterings between electrons
and electrons, or between positrons and positrons, do not change the distribution function in a one-dimensional plasma.

The ohmic heating rate per unit volume is
\be
\dot u_{\rm ohm} = {J_\parallel^2\over  \sigma_{\rm cond}} = 4(\beta\gamma)^3 \tau_{\rm T,min}^2 {\sigma_{+-}\over\sigma_{\rm T}}
     {m_ec^3\over \sigma_{\rm T} R^2}\left[{n_\pm(r)\over n_\pm(R)}\right]^2.
\ee
The total dissipation rate depends on the density profile $n_\pm(r)$.  In the absence of collisions, $n_\pm(r) \propto B(r)$
given that the pairs are confined to the lowest Landau level but warm enough to overcome the strong surface gravity of the magnetar.
Here, we are considering the opposite, strongly collisional regime.  Although the pressure gradient in a collisional plasma
acts to reduce a density gradient, the collisions here are are dominated by pair annihilation.   The density distribution then is
determined by local energy and annihilation equilibrium, not by a force balance in the direction parallel to ${\bf B}$.
We show in Section \ref{s:eqdepth} that this equilibrium corresponds to $n_\pm(r) \propto n_{\pm,\rm min}(r) \propto B(r)$,
the same scaling as in the collisionless case.

An analytic approximation to the volume integral of $\dot u_{\rm ohm}$ is now available.
Substituting $\sigma_\pm = \sigma_\pm(s) \propto B^{-1}$ (neglecting the exponential factor in Equation (\ref{eq:sigpm2}))
and integrating over the volume of a flux bundle of
cross-sectional area $A_\perp(r) = A_\perp(R) B(R)/B(r)$, we find that the ohmic energy input is
uniform per unit length of the bundle.  Normalizing the dissipated power to the surface
cross sectional area $A_\perp(R)$ of the bundle, we obtain (Figure \ref{fig:ohm})
\ba\label{eq:edotohm}
   {\dot E_{\rm ohm}\over A_\perp(R)} &\;\simeq\;& l\cdot \dot u_{\rm ohm}(R) \nn
   &=& 1.1\times 10^{24} \beta^2\gamma \left({l\over R}\right) {\tau_{\rm T, min}^2\over R_6 B(R)_{15}}.
\ea
Here $l$ is the length of the bundle.  

Note that this result does not depend explicitly on the magnetospheric pair density $n_\pm$, but instead on the minimum
density needed to supply the current.  
The heating rate in Equation (\ref{eq:edotohm}) is only a lower limit as it represents the effects of collisional
resistivity.

\subsection{Anomalous Resistivity in the Electron-Ion Atmosphere}\label{s:anom}

A promising site for the excitation of plasma turbulence is the transition layer separating
the magnetosphere from the much thinner magnetar atmosphere, which is dominated by electrons and ions.
Here, the ion (proton) density rises above the magnetospheric positron density.
We now consider the stability of the current-carrying plasma in this layer, for two reasons.
First, it is a promising source of coherent plasma emission in the optical-IR band (Section \ref{s:optical}).
And, second, scattering off the turbulence may slow down the drift rate of positrons into the atmosphere.

The atmosphere supports ion acoustic modes with a characteristic speed
\ba
\left({\omega\over k}\right)^2 &=& c_s^2\left(T_e, n_+/n_p\right) = {\omega_{Pi}^2\over \omega_{Pe}^2 + \omega_{P\pm}^2}{T_e\over m_e}\nn
&=& {T_e\over m_p}\left(1 + {2n_+\over n_p}\right)^{-1}
\ea
and frequency $\omega \lesssim \omega_{Pi}$.
Here we include the screening effects of both the neutralizing electrons and downward-drifting pairs,
with $\omega_{Px} = (4\pi n_x e^2/m_x)^{1/2}$ representing the plasma frequency of species $x$.  We also assume that the electrons are heated
collisionlessly to $T_e \sim \beta^2 m_ec^2$ at the top of the atmosphere (as appropriate for a one-dimensional plasma).  The ion sound speed is reduced
compared with the pair-free result, although the mode remains weakly damped only as long as $T_e/T_p > 1+2n_+/n_p$.

Ion acoustic modes are linearly unstable when the electron drift speed through the atmosphere exceeds $c_s$
and when the ion temperature $T_p \ll T_e$ (e.g. \citealt{kulsrud05}),
\be
\beta_{\rm dr} = {n_{\pm,\rm min}(R)\over n_e + n_\pm}\beta = {n_{\pm,\rm min}(R)\over n_p + 2n_+}\beta  > {c_s\over c}.
\ee
This condition is easily satisfied even in the upper parts of the ion-dominated atmosphere,
where $n_p$ rises above $n_+ = n_\pm/2$.  Estimating $\beta \sim 0.6$, $T_e \sim 0.4\,m_ec^2$
(a characteristic result for the equilibrium magnetospheric temperature; Section \ref{s:results}), we find
$c_s/c \sim 0.015$.  The ion acoustic mode is unstable as long as
\be
n_p < 40\,n_{\pm,\rm min}(R) \sim 6 n_+(R).
\ee

An essential point here is that ion acoustic turbulence does not feed back on the magnitude of the current, which is imposed by
the rigid background magnetic shear.  This means that the unstable mode must grow either until limited by nonlinear damping,
or until the ions are sufficiently heated that the mode is linearly damped.  Ion heating is, however, limited by mutual
Coulomb scattering into higher Landau states, followed by rapid radiative deexcitation.  Assuming cool ions,
there is a powerful continuing exchange of energy between particles and waves.  The scattering time of $e^\pm$
off modes of frequency $\omega$ cannot be shorter than $\tau \sim \omega^{-1}$.  Hence the heating rate due to an anomalous
resistivity ($\sigma_{\rm anom} = n_\pm e^2 \tau/\gamma^3 m_e$) is bounded above by
\be
\dot u_{\rm anom} \sim {J_\parallel^2\over\sigma_{\rm anom}} < \left({n_{\pm,\rm min}\over n_\pm}\right)^2
\cdot n_\pm m_ec^2\beta^2\gamma^3\omega.
\ee

The heating rate per unit area is obtained by integrating $\dot u_{\rm anom}$
over the scale height $h = (T_e/m_p g)(1+2n_+/n_p)$ of the transition layer (here $g$ is surface gravity).
The maximal value of this heat flux exceeds the kinetic energy flux of downward-moving pairs, which is
${1\over 4}n_\pm(R) m_e (\beta c)^3$ in a non-relativistic approximation,
by some four orders of magnitude.  If we adjust the scattering time $\tau$ so that these two quantities are comparable, then
\be
\tau \sim 4 {h\over \beta c}\left({\beta_{\rm dr}\over \beta}\right)^2.
\ee
The diffusion coefficient of pairs moving parallel to ${\bf B}$ through the transition layer is $D \sim (\beta c)^2 \tau$.
Given that the positron density at the base of the layer is much smaller than the magnetospheric density $n_+(R)$,
due to rapid annihilations with atmospheric electrons, there is a downward diffusive flux of positrons.
This downward flux is suppressed compared with the first estimate $F = {1\over 2} n_+(R) \beta c$,
\be
F  \sim  {D\over h} n_+(R)  =  {1\over 2}n_+(R) \beta c \cdot 8\left({\beta_{\rm dr}\over \beta}\right)^2.
\ee
The suppression factor is smaller than 0.1 for $\beta_{\rm dr} < 0.1\,\beta$.
The annihilation rate of pairs in the atmosphere may therefore be substantially reduced.

There is a useful contrast here with the relativistic double layer model of \cite{BT07}.  There the downward-moving
electrons or positrons have relativistic energies, $\gamma \sim 10^3$, and would not easily be repelled by a turbulent
transition layer.

\section{Photon Emission and Pair Equilibrium}\label{s:results}

We now consider the radiative emission from a pair-loaded and current-carrying magnetic flux bundle
anchored in the crust of a magnetar, and the corresponding evolution of the temperature and density of
the pair plasma.   We start with an analytic estimate of the volume-integrated
photon luminosity sourced by $e^+$-$e^-$ annihilation.  This is then compared with the thermal radiative flux
driven by the annihilation of positrons on the stellar surface and with the ohmic heating rate derived in Section \ref{s:current}.
The equilibrium state in which radiative cooling balances collisional heating is shown to have a much larger pair
density than the minimum needed to supply the magnetospheric current.

In the second part of this Section, we present results of the MC simulations of the escaping annihilation radiation and pair creation
within the plasma, for a range of scattering depths and thermal speeds.  We then show that the current-carrying plasma so described
can relax to state of combined energy and annihilation equilibrium, and that this equilibrium state is stable.

\subsection{Luminosity}

Consider the annihilation-driven photon emission from a slender, closed magnetic arcade (Figure \ref{fig:fluxtube}).
We start with the cross section (\ref{eq:sigma_tot}) for photon emission by pair annihilation, using the
soft-photon approximation implied by Equation (\ref{eq:lowfreq}).
The MC results show that, to a good approximation, the escaping photon spectrum has this slope up to a maximum
energy $\hbar\omega \sim m_ec^2$.  Hence, the energy release per unit volume and time is
\ba\label{eq:udotann}
         {d\dot u_{\rm ann}\over d\ln\omega}
         &=& 2\beta c \cdot 2\hbar\omega{d\sigma_{\rm ann}\over d\ln\omega}\cdot {n_\pm^2\over 8}\nn
         &=& \hbar\omega\,{2\pi r_e^2 c\over B/\BQ}f(\beta)\,n_\pm^2,
\ea
where $f(\beta)$ is defined in Equation (\ref{eq:lowfreq}).  (A typical value is $f(0.6) = 0.36$.)
Integrating over volume, the emission is constant per unit length of the arcade for the same reason
that the ohmic heat input is constant, and
\ba\label{eq:edotvol}
         {1\over 2\,A_\perp(R)}{d\dot E_{\rm ann}\over d\ln\omega} &\;\simeq\;& {l\over 2}\cdot{d\dot u_{\rm ann}\over d\ln\omega}\biggr|_R \nn
         &=& {3\hbar \omega\over 2B(R)/\BQ} {\tau_{\rm T}^2(R) c\over \sigma_{\rm T}R } \left({l\over R}\right) f(\beta).\nn
\ea
Here we have substituted for the optical depth (\ref{eq:tauTR}) along the active flux bundle.
The factor ${1\over 2}$ on the left-hand side counts the area of both magnetic footpoints.
In this case, $l$ is not the total length of the arcade, but the length of that portion with magnetic flux density
exceeding $B \sim 5\,\BQ$.  
Numerically,
\ba\label{eq:edotvol2}
   &&{1\over 2\,A_\perp(R)}{d\dot E_{\rm ann}\over d\ln\omega} = 0.5\times 10^{23}\,{\tau_{\rm T}^2(R)_1\over R_6 B(R)_{15}}\nn
 &&  \quad\quad \times \left({\hbar\omega\over 100~{\rm keV}}\right)\left({l\over R}\right)f(\beta)\quad {\rm erg~cm^{-2}~s^{-1}}.\nn
\ea

A quasi-thermal (blackbody) component of the spectrum will be sourced by the impact and annihilation of pairs
on the magnetar surface.  This is modeled in detail by \cite{KT20}, who finds that the conversion efficiency of annihilation energy
to blackbody photons is $\varepsilon_{\rm bb} \sim 0.2-0.3$ when the pairs are transrelativistic.
 The pairs annihilate at a shallow depth in the atmosphere because they backscatter off cold atmospheric
  electrons with the large cross section (\ref{eq:sigpm2}), forming a cold positron gas. Most annihilations occur
  within this cold gas.  Most of the energy from each annihilation is carried by a single hard photon that propagates
  nearly perpendicular to ${\bf B}$ and so loses a small fraction
  of its energy to electron recoil, after which it experiences a much reduced scattering depth.  A modest fraction
  of the total energy of the downflowing pairs is conducted deep into the atmosphere, where it is thermalized.
The main uncertainty here is the flux of positrons passing
through the transition layer between the ion-dominated atmosphere and the pair-dominated magnetosphere.  The
downward flux is ${1\over 2}\beta c\, n_+(R)$ in the absence of ion-acoustic turbulence, but is plausibly reduced
by an order of magnitude by turbulent scattering (Section \ref{s:anom}).

We can compare this thermal output with the volumetric output in Equation (\ref{eq:edotvol}).
Normalized to the area of the heated surface, the total blackbody luminosity $\dot E_{\rm bb}$ is given by
\ba\label{eq:edotsurf}
         {\dot E_{\rm bb}\over 2\,A_\perp(R)} &=& \varepsilon_{\rm bb}\cdot \beta {n_+(R)\over 2}2m_ec^3  \nn
         &=& \varepsilon_{\rm bb}\cdot \beta \tau_{\rm T}(R) {m_ec^3\over\sigma_{\rm T}R}.
\ea
Numerically,
\be\label{eq:edotsurf2}
{\dot E_{\rm bb}\over 2\,A_\perp(R)} = 0.9\times 10^{23}\,\beta\left({\varepsilon_{\rm bb}\over 0.25}\right){\tau_{\rm T}(R)_1\over R_6}\quad {\rm erg~cm^{-2}~s^{-1}}.
\ee
   The corresponding blackbody (effective) temperature, assuming radiation into a single polarization mode (the O-mode) is
     $T_{\rm bb} = 0.65\,[\beta(\varepsilon_{\rm bb}/0.25)\tau_{\rm T}(R)_1/R_6]^{1/4}$ keV.
Comparing this with Equation (\ref{eq:edotvol}) shows that the output at 100 keV competes with the O-mode
black-body peak when $\tau_{\rm T}(R) \gtrsim 10 (\varepsilon_{\rm bb}/0.25)^{-1}$.  

To obtain the total hard X-ray and blackbody luminosity of a magnetar, the `fluxes' in Equations (\ref{eq:edotvol}) and
(\ref{eq:edotsurf}) must be multiplied by the total cross section $2A_\perp(R)$ of the active flux bundle at the magnetar surface.
The relative strength of these two components is independent of the shape of the magnetic arcade, e.g., whether it has
a locally circular cross section, or a sheet-like shape as in Figure \ref{fig:fluxtube}.
One obtains $\sim 10^{35}$ erg s$^{-1}$ when $\tau_{\rm T}(R) \sim 10$ and the heated area is $\sim 10^{12}$ cm$^2$,
about 10 percent of the surface area of a neutron star.

\begin{figure*}
  \epsscale{0.9}
  \plottwo{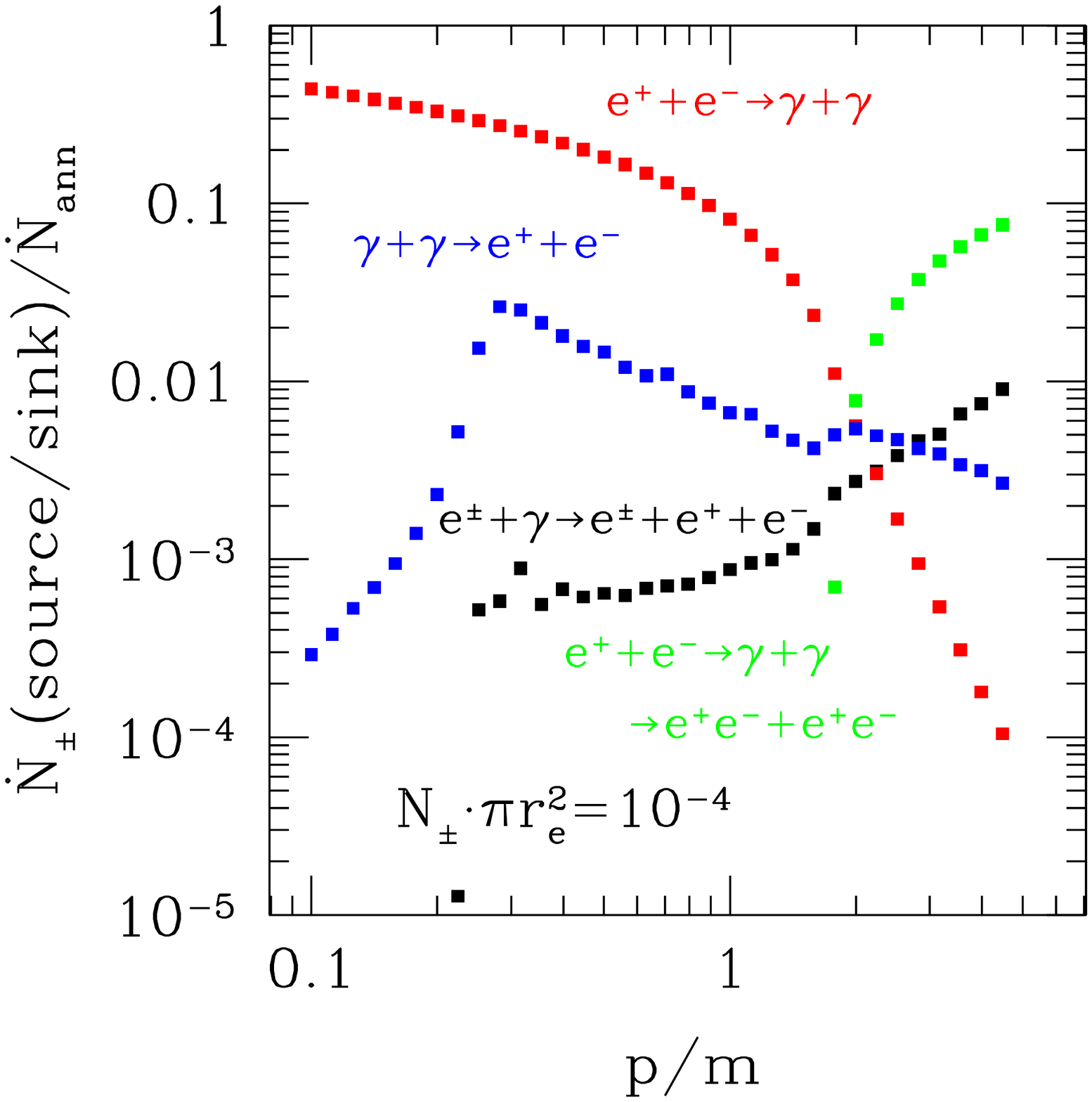}{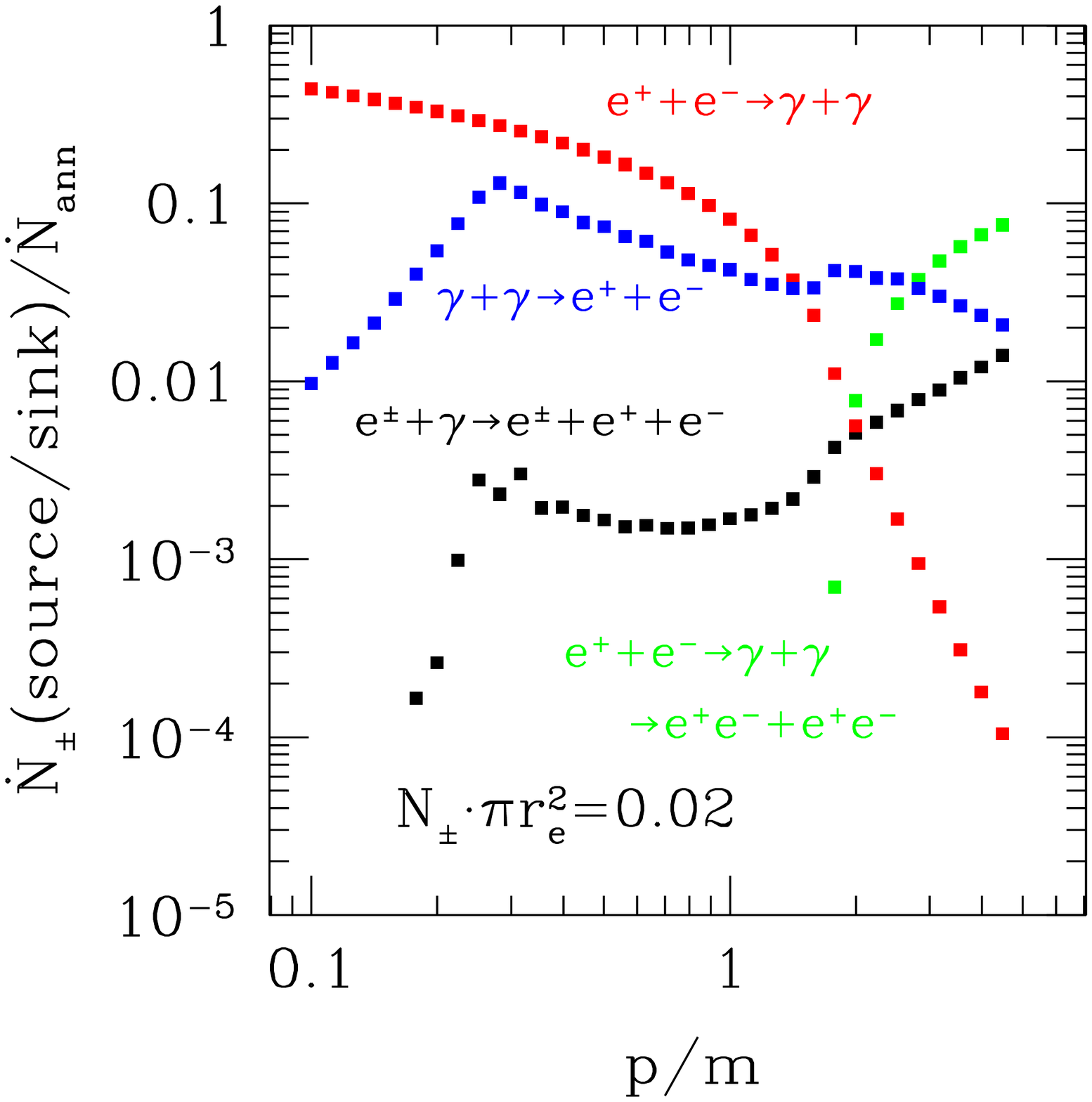}
  \vskip -1in
  \plottwo{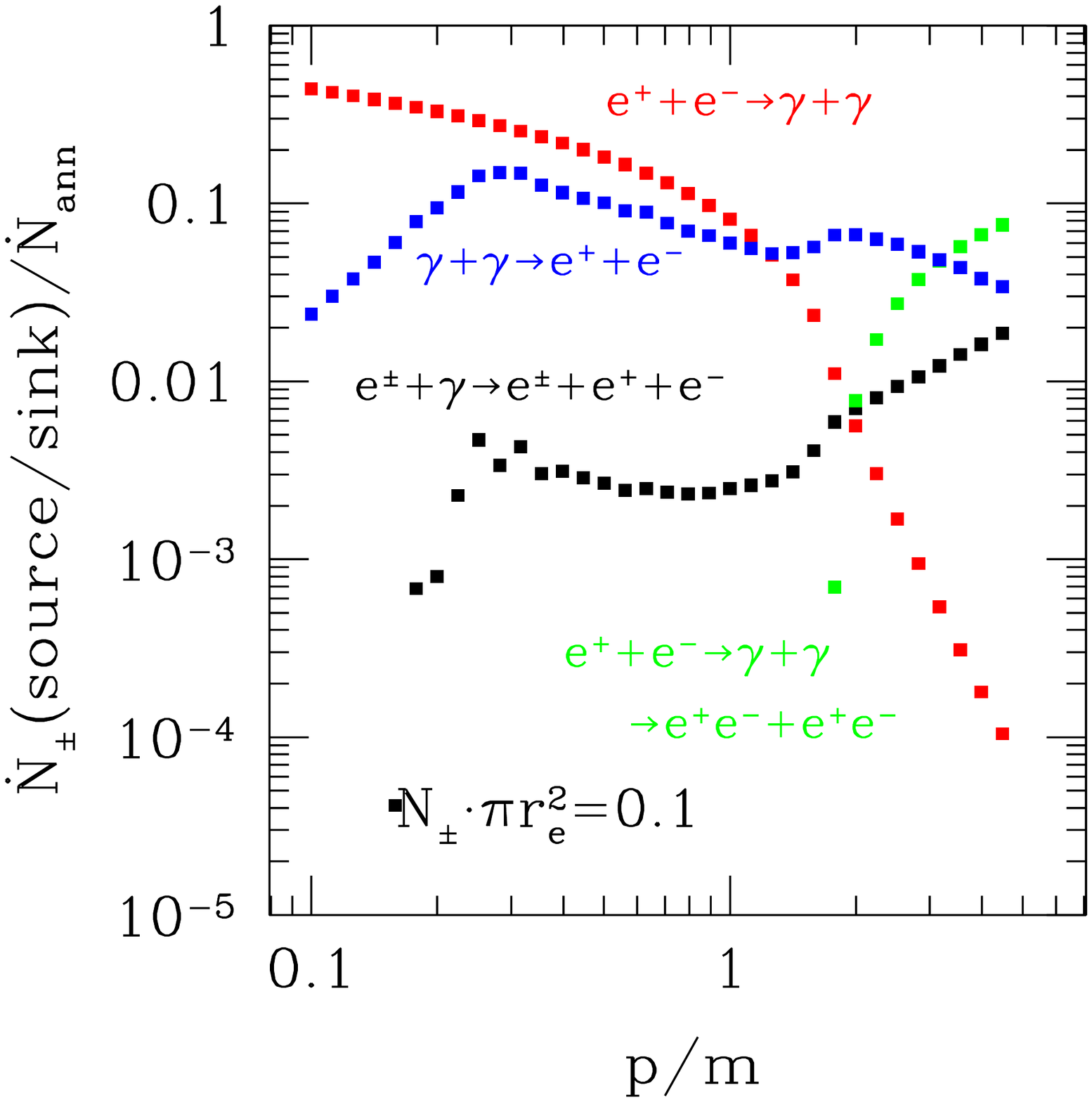}{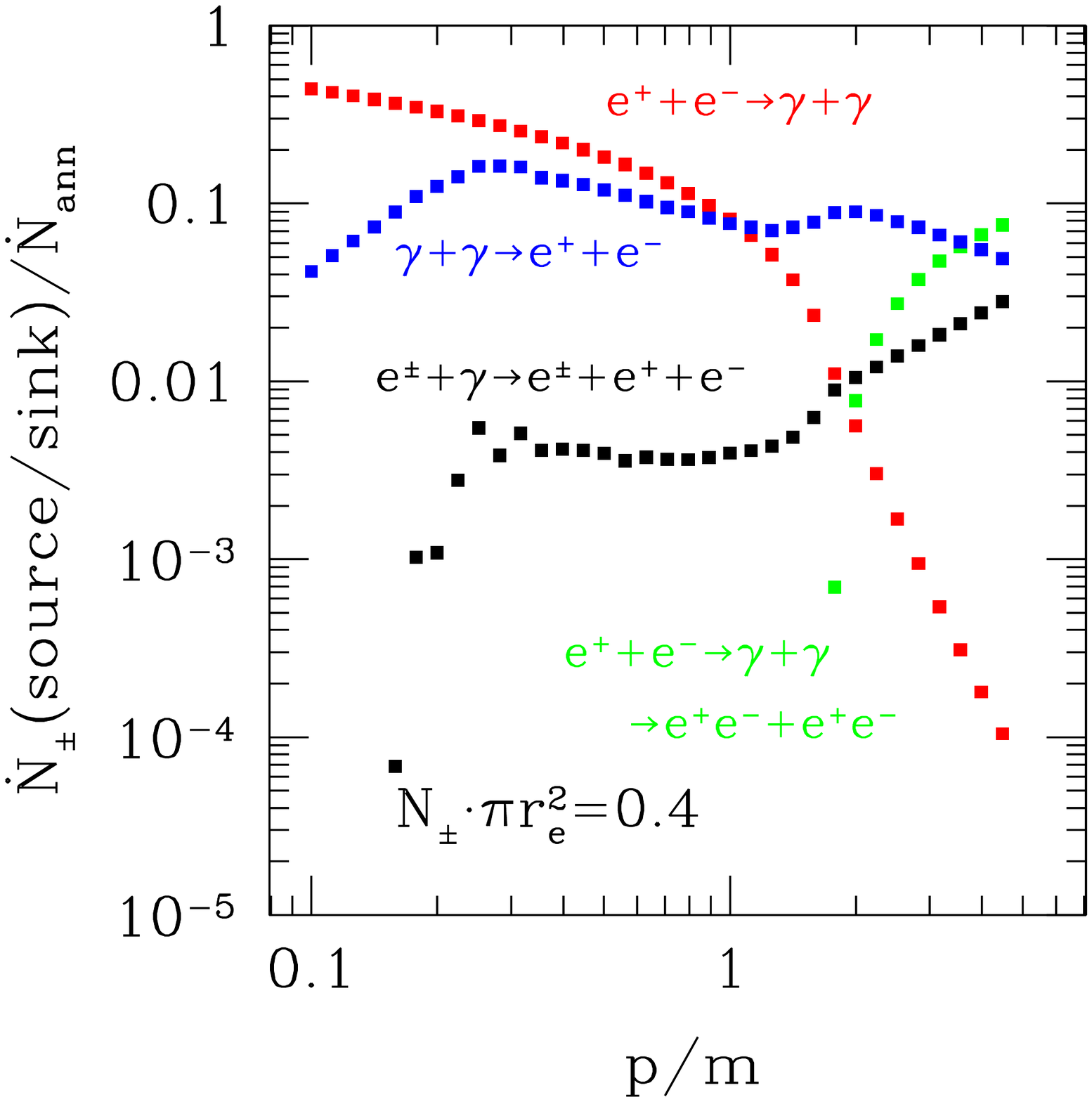}
    \vskip -1in
\plottwo{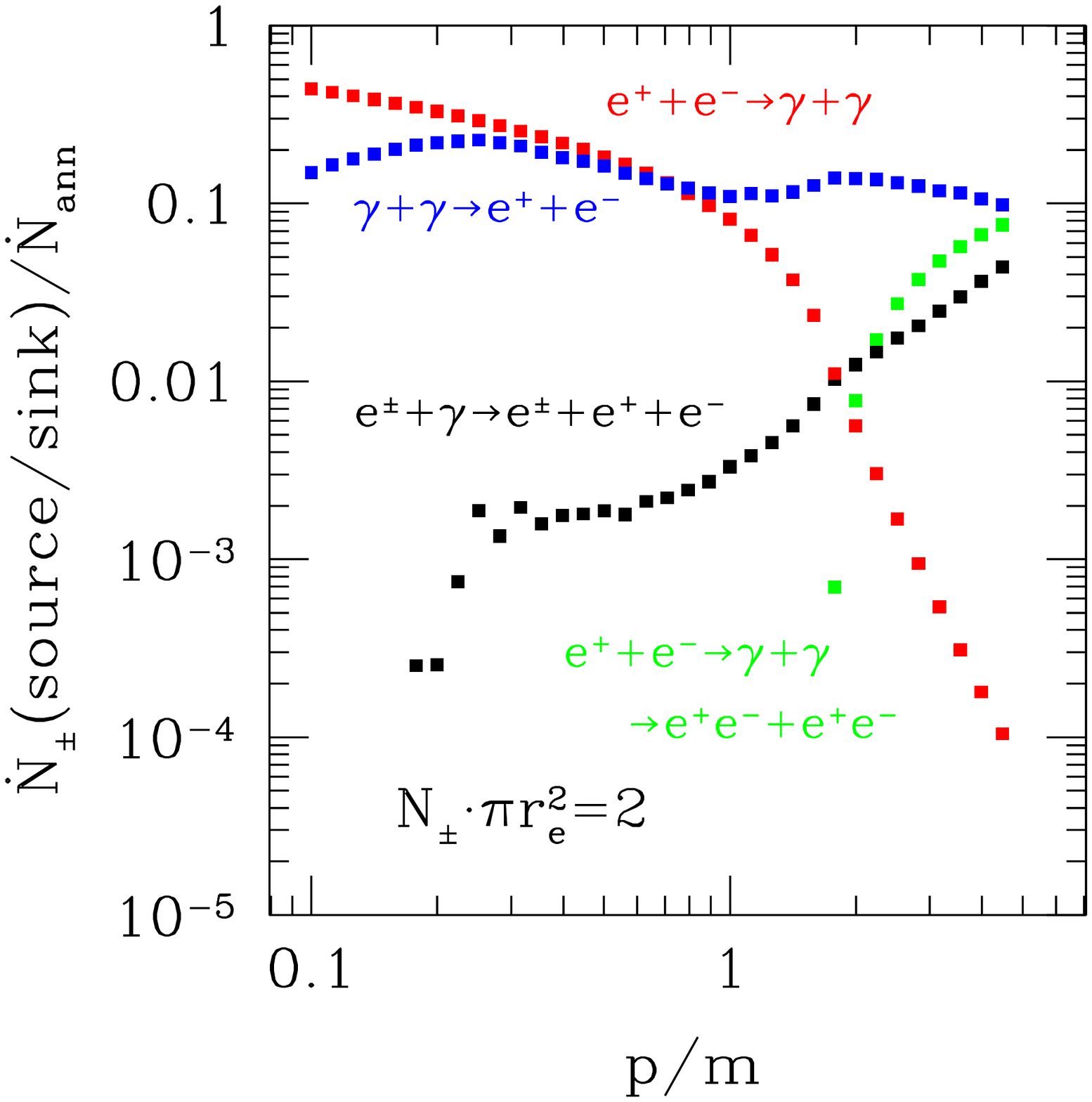}{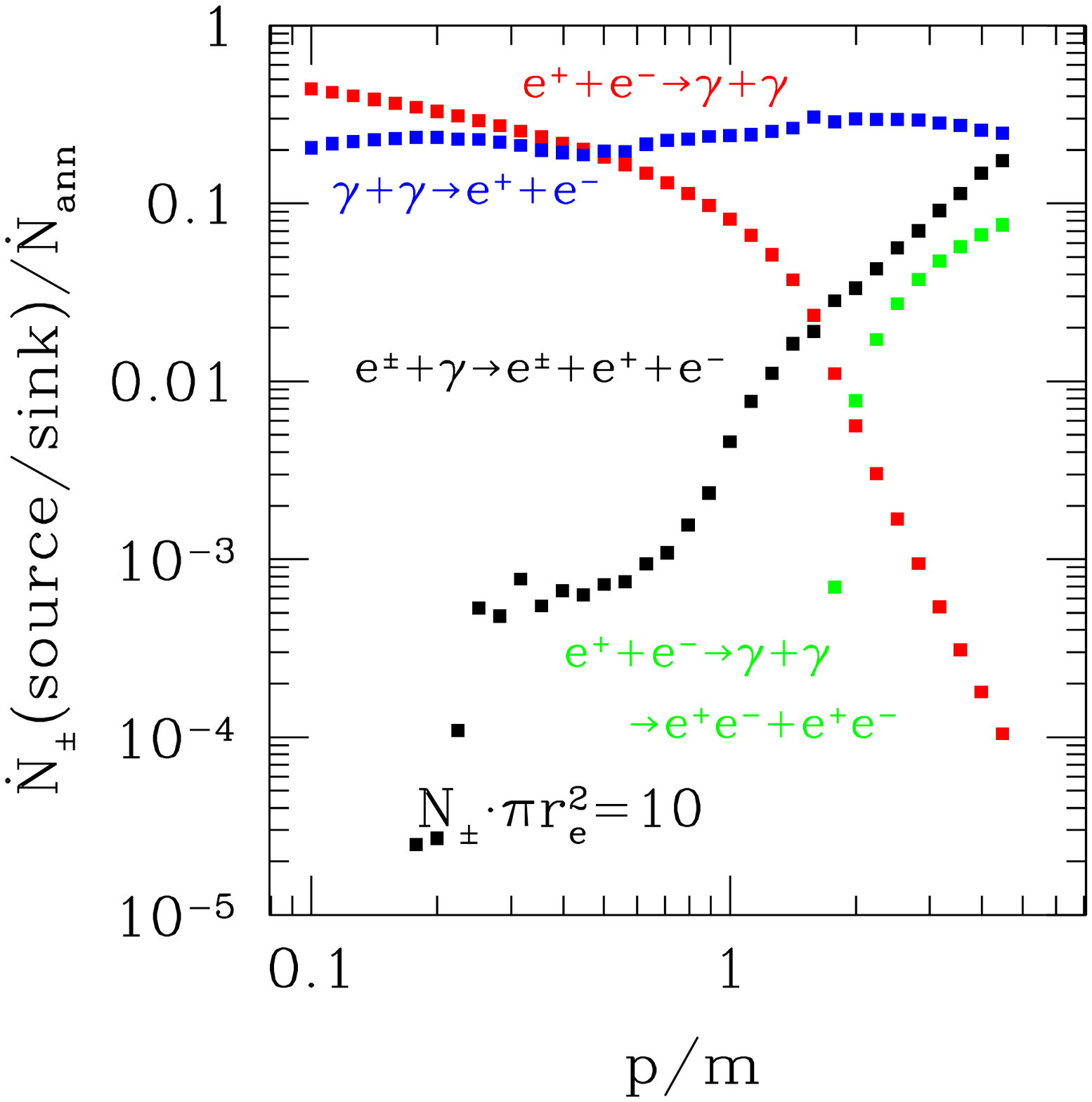}
\vskip -0.7in
\caption{Yield of pairs in a magnetized pair plasma ($B = 10\,\BQ$) versus thermal momenta $\pm p$,
  normalized by the total number of two-photon annihilation events (including all final states).
  Black, blue, green points show processes with positive yield;  red with negative yield.
  From upper left to lower right:  $N_\pm \pi r_e^2 = 10^{-4}$, 0.02, 0.1, 0.4, 2, 10.
  Black points:  a photon scatters off an electron or positron into a state that
  converts directly to a magnetized pair, $\gamma + e^\pm \rightarrow \gamma + e^\pm \rightarrow e^+ + e^- + e^\pm$.
  Blue points:  pair creation by photon collisions, $\gamma + \gamma \rightarrow e^+ + e^-$.
  Green points:  annihilation of a pair into two photons, each of which is above threshold for
  direct pair conversion, $e^ + e^- \rightarrow \gamma + \gamma \rightarrow e^ + + e^- + e^+ + e^-$.
  Red points:  annihilation into two photons, neither of which is above the threshold for direct
  pair conversion.}
\end{figure*}\label{fig:pairrates}

\begin{figure}[t]
  \epsscale{1.2}
  \vskip -0.3in
\plotone{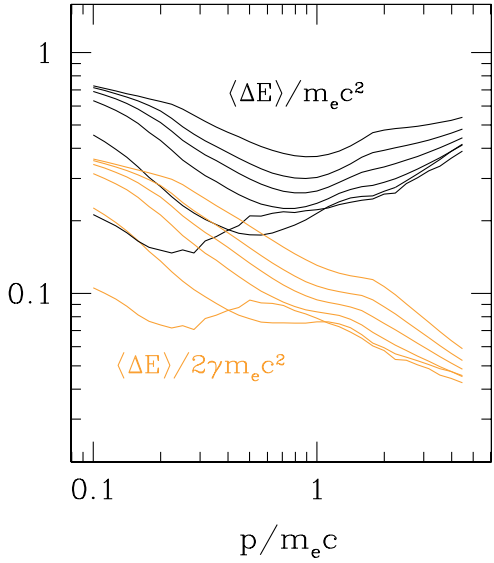}
\vskip -0.8in
\caption{Black curves:  mean energy released (in the form of escaping photons) per two-photon annihilation event in
  the magnetospheric plasma.  Orange curves:  a fraction of the total energy of each annihilating pair.
  From bottom to top (on the left):  $N_\pm \pi r_e^2 = 10^{-4}$, 0.02, 0.1, 0.4, 2, 10.}
\end{figure}\label{fig:radenergy}

\begin{figure}
  \epsscale{1.2}
  \vskip -0.3in
\plotone{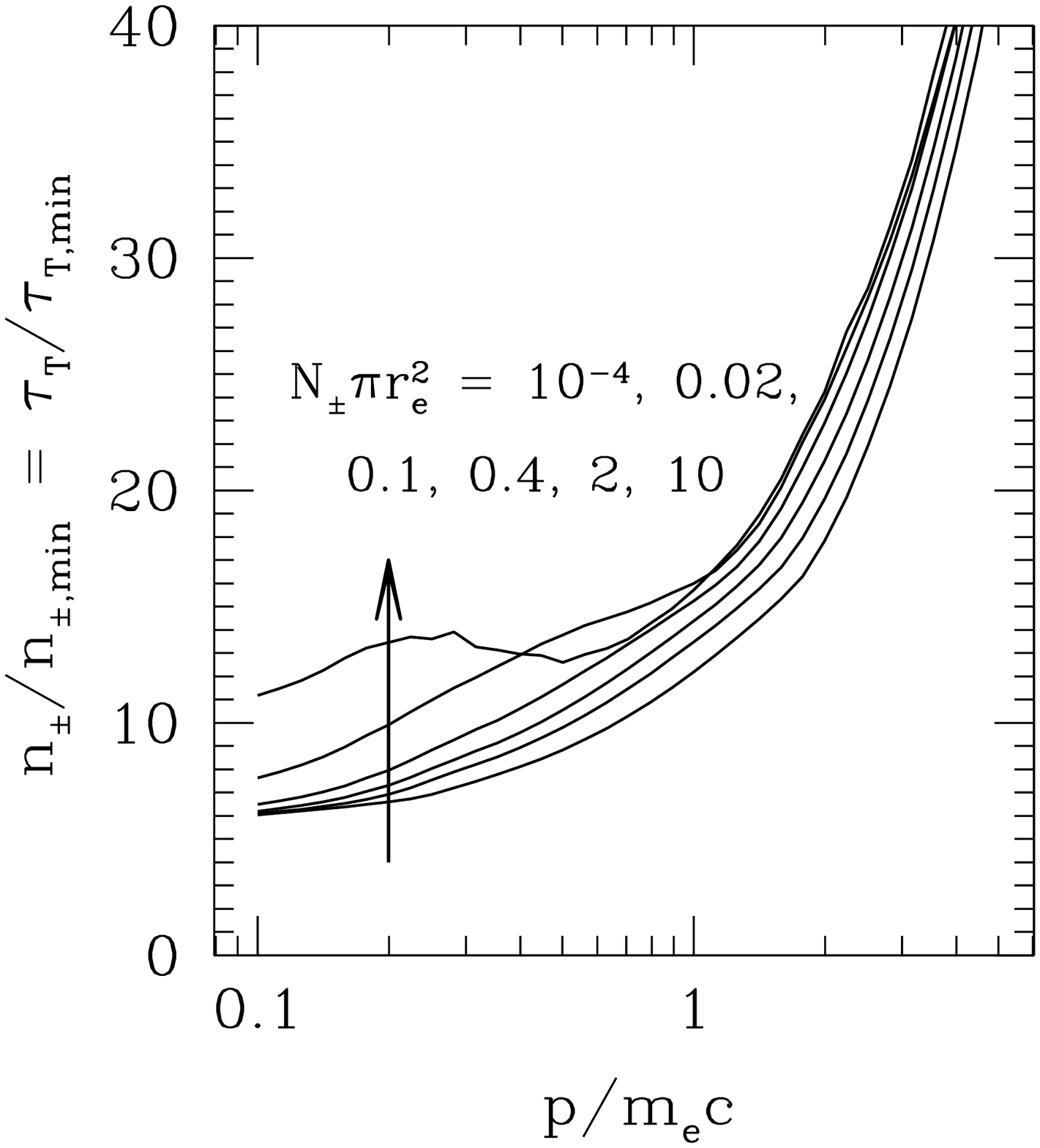}
\vskip -0.8in
\caption{Pair density within a pair-loaded, current-carrying magnetic flux
  bundle which is in a state of balance between collisional heating and radiative cooling.  This equilibrium implies a particular
  ratio between $n_\pm$ and the density corresponding 
  to a marginally charge-starved state ($n_{\pm,\rm min} = |J|/\beta ec$), which varies
  mildly with the thermal momentum of the pairs.  Below this optical depth, ohmic heating overwhelms radiative energy
  loss, raising the temperature of the pairs.  According to Figure \ref{fig:pairrates}, this then forces an excess of
  internal pair creation over annihilation.  The equilibrium state corresponds to a balance both between heating
  and cooling, and pair creation and annihilation.  }
\end{figure}\label{fig:ohm2}

\subsection{Equilibrium Scattering Depth}\label{s:eqdepth}

The hard X-ray output in Equation (\ref{eq:edotvol}) is able to compensate the collisional dissipation in Equation
(\ref{eq:edotohm}) only if the magnetospheric pair density is much greater than the minimum needed to supply the current.
This result is independent of the surface magnetic field:
\ba\label{eq:taueq}
   {\tau_{\rm T}\over\tau_{\rm T,min}} &=& \left({\pi\over 2\alpha_{\rm em}}\right)^{1/2}
   \left({\hbar\omega_{\rm max}\over m_ec^2}\right)^{-1/2} {\beta\gamma^{1/2}\over f(\beta)^{1/2}}\nn
    &=& 15\,\left({\hbar\omega_{\rm max}\over m_ec^2}\right)^{-1/2}
   {\beta\gamma^{1/2}\over f(\beta)^{1/2}}.
\ea
Here we approximate the escaping luminosity in terms of the emitted spectrum
given by Equation (\ref{eq:edotvol}) up to a maximum frequency $\hbar\omega_{\rm max} \sim m_ec^2$.  (The numerical results
for the emitted spectrum presented in Section \ref{s:spectrum} show this to be a good approximation.)

An even greater pair density and optical depth is required to balance resistive heating when the resistivity is enhanced by
a collective instability of the plasma.  

We conclude that a trans-relativistic and collisional state is available to the inner parts of the magnetar
magnetosphere.  In energy equilibrium, the plasma
is dense enough to support the current with a drift speed much smaller than the $e^\pm$ thermal speed,
$\beta_{\rm dr}/\beta \lesssim 0.1$.  Given that the $e^\pm$ backscattering cross section is
$\sigma_{+-} \sim 7\,B_{15}^{-1}(p/m_ec)^{-1}\sigma_{\rm T}$ for $0.3 B_{15}^{1/3}m_ec \lesssim p \lesssim m_ec$ (Figure \ref{fig:sigma_pp}),
collisionality is maintained for
\be
\tau_{\rm T,min} \gtrsim (0.02-0.03)\,B_{15}.
\ee

This evaluation of the equilibrium pair density assumes that annihilation in the magnetosphere is faster than
the annihilation of positrons impacting the magnetar surface.   To show that this is the case, we integrate
the volumetric annihilation rate into two photons, both below the pair conversion threshold, $\dot n_{\rm ann}(2\gamma)
= 2\beta c \,\sigma_{\rm ann}(2\gamma) n_\pm^2/8$, where $\sigma_{\rm ann}(2\gamma) \simeq 1.7 (\BQ/B) \sigma_T$
(Figure \ref{fig:sigma_ann}).  As before, this is distributed uniformly along the active flux bundle,
\ba\label{eq:ndotsurf}
\dot N_{\rm mag} &=& \int \dot n_{\rm ann}(2\gamma) dV \nn
&\simeq& {1.7\,\beta\over B(R)/\BQ} {\tau_{\rm T}^2(R)c\over\sigma_{\rm T}R}\left({l\over R}\right)A_\perp(R).
\ea

The surface annihilation rate at both end of the flux bundle is $\dot N_{\rm surf} = \varepsilon_{\rm ann}'\,\beta c\cdot {1\over 2}n_+(R) \cdot 2A_\perp(R)$.
We assume that a downward-moving positron annihilates a fraction $\varepsilon_{\rm ann}'$ of the time, rather than backscattering.
Comparing with the volumetric annihilation rate, one finds
\be
   {\dot N_{\rm mag}\over \dot N_{\rm surf}} =  {1.5\over (\varepsilon_{\rm ann}'/0.5)}{\tau_{\rm T}(R)_1\over B(R)_{15}}\left({l\over R}\right).
\ee
When this ratio is larger than unity, pair creation and annihilation in the magnetosphere can be treated in local competition.
However, when $\dot N_{\rm mag}$ falls below $\dot N_{\rm surf}$, the collisional equilibrium state for the pair plasma
becomes inconsistent.  Alternative states include a more dilute, transrelativistic plasma sourced nonlocally by
photon collisions (Section \ref{s:nonlocal}), or the relativistic double layer found by \cite{BT07}.

\subsection{Pair Creation:  Numerical Results}

Here and in Section \ref{s:spectrum} we present results obtained from the MC code that was described in Section \ref{s:monte}.
The background magnetic field is taken to be $B = 10\,\BQ$.
The relative rates of various channels for pair creation and destruction are shown in Figure \ref{fig:pairrates},
all normalized to the total annihilation rate
\be\label{eq:dotnann}
\dot n_{\rm ann} = 2\beta c \Bigl[\sigma_{\rm ann}(2\gamma) + \sigma_{\rm ann}({\rm 1\gamma/1p}) +
\sigma_{\rm ann}({\rm 2p})\Bigr]{n_\pm^2\over 8}.
\ee
This comparison is made for a range 
of scattering depths through the pair plasma ($N_\pm \pi r_e^2 = 10^{-4}-10$) and a range of thermal momenta $p$
(extending from $p = 0.1\,m_ec$ to the threshold for scattering into the first excited Landau level, $(2B/B_Q)^{1/2} m_ec$).
The red points show the primary pair sink through annihilation into two photons, both
of which are below the pair conversion threshold.

The dominant source of pairs is photon collisions (blue points in Figure \ref{fig:pairrates}), which grows in strength
with the scattering depth across the plasma.  That is because (at low to moderate scattering depth)
the density of photons scales as $n_\gamma \propto n_\pm^2 \sigma_{\rm ann} \propto \tau_{\rm T}^2 / B$.  The optical depth to
photon collisions is proportional to $n_\gamma \sigma_{\gamma\gamma} \propto n_\gamma \, B \propto \tau_{\rm T}^2 \,B^0$.  Results
are shown for a magnetic field $10\,\BQ$, but this scaling argument shows that they are only weakly dependent
on $B$ as long as $B \gtrsim 5\,\BQ$ and our large-$B$ approximation to the annihilation spectrum is valid.

The rates of pair creation and annihilation are shown to cross at $p > m_ec$ when the scattering depth is small,
settling down to $p\sim (0.5-1)m_ec$ when $N_\pm \pi r_e^2 \gtrsim 1$.  
Pair creation occurs at a higher rate for thermal momenta above this bound, implying that the plasma experiences a reduction
in the mean energy per charge in the absence of radiative losses and moves to a state of annihilation equilibrium (Section \ref{s:equil}).

Tracking the absorption and scattering of successive photons allows us to evaluate the radiative energy loss
per annihilation event, show in Figure \ref{fig:radenergy}.  This allows us to refine the calculation of the balance
between ohmic heating in Equation (\ref{eq:taueq}), as shown in Figure \ref{fig:ohm2}.  The equilibrium scattering
depth is about 10 times the minimum value $\tau_{\rm T,min}$ that will support the current, in agreement with the analytic
estimate in Equation (\ref{eq:taueq}).

\begin{figure*}
  \epsscale{1.1}
  \vskip -0.3in
\plottwo{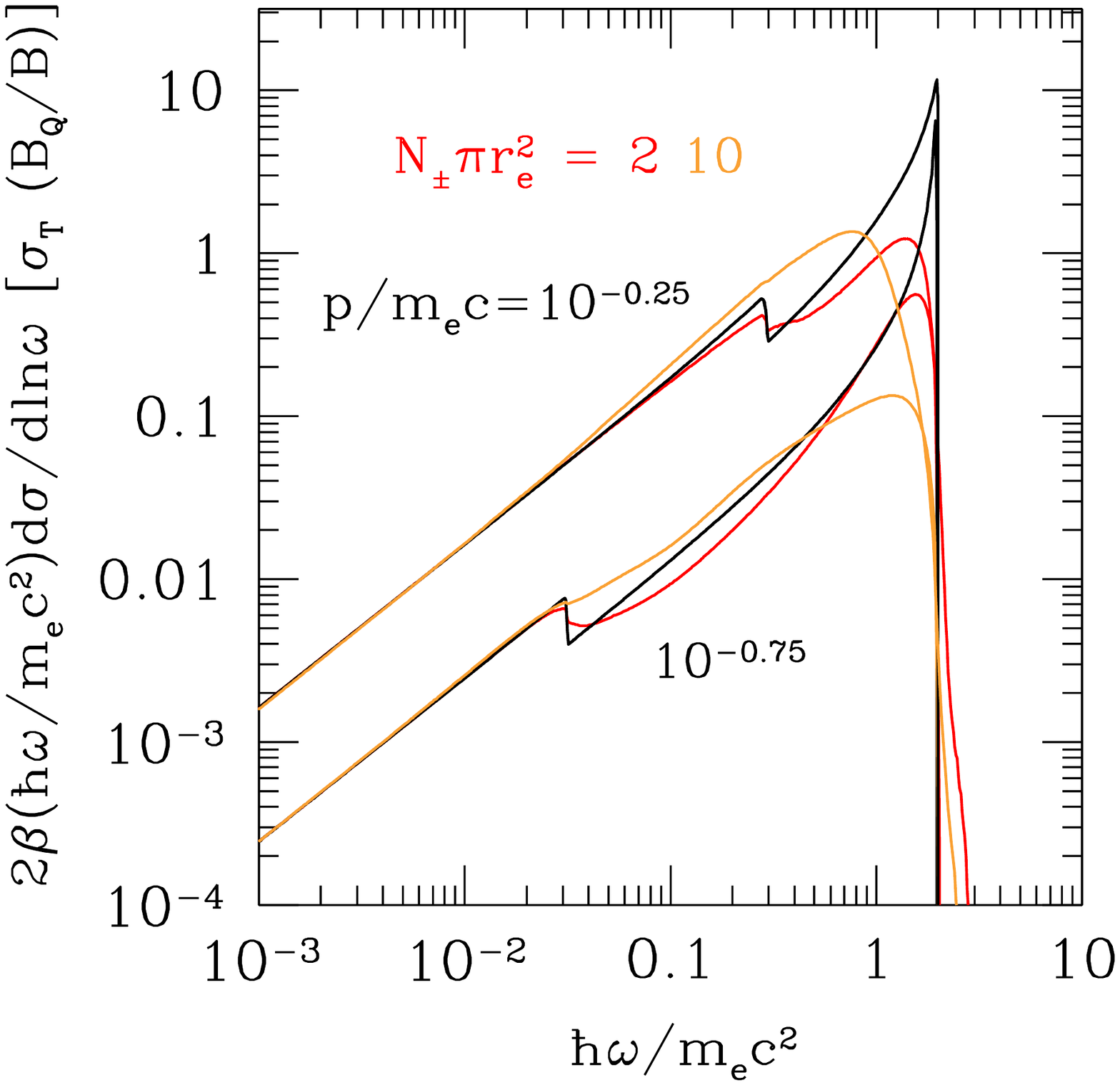}{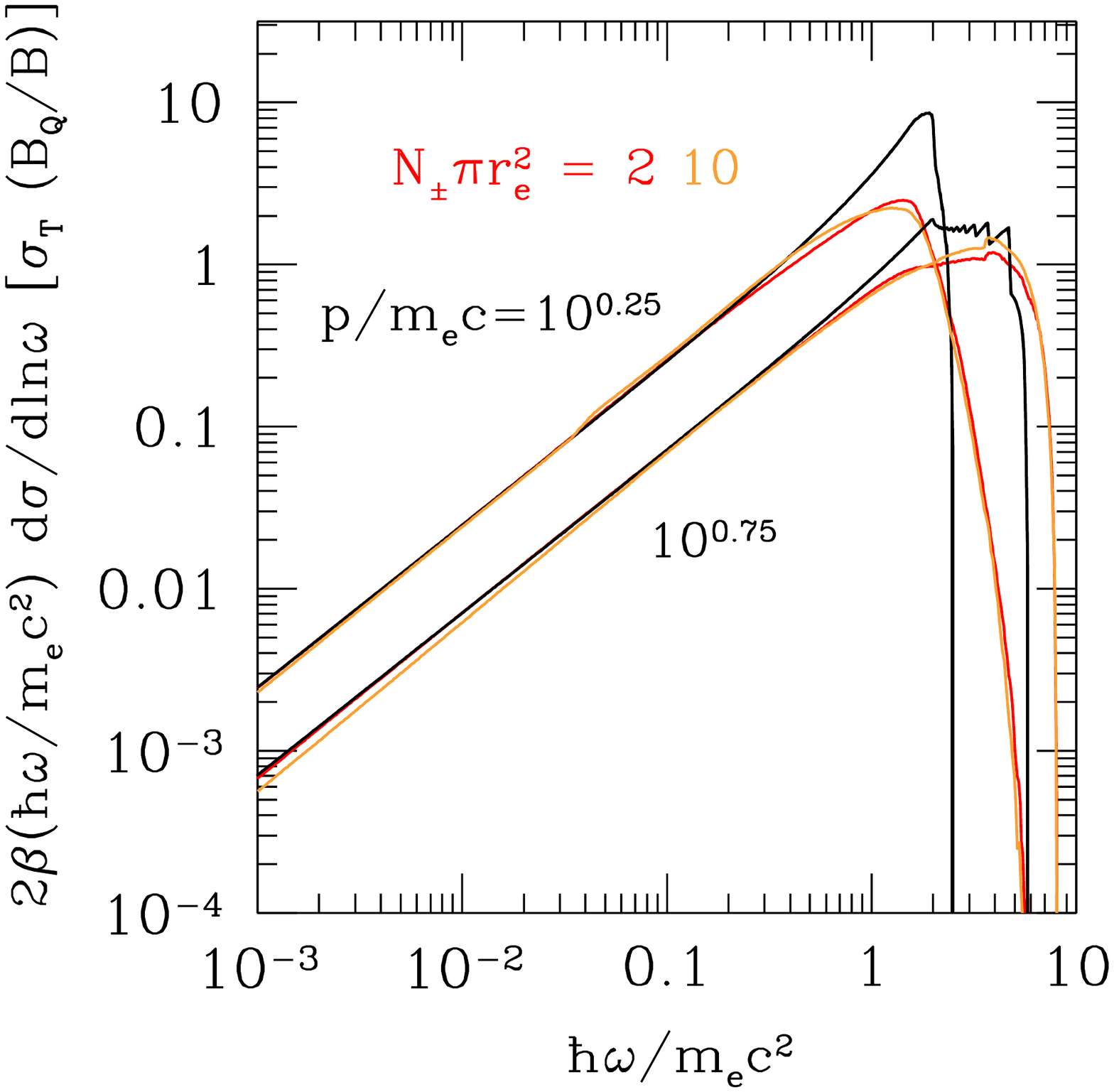}
\vskip -0.7in
\caption{Sample of output energy spectra (red and orange curves) for the models with $B = 10\,\BQ$, transverse scattering depth $N_\pm \pi r_e^2 = 2$,
  and 10 and a range of $e^\pm$ thermal momenta $p/m_ec$.   Black curves show for comparison the emission spectrum before the effects
  of scattering and photon collisions are taken into account, as given
  by Equation (\ref{eq:sigma_tot}).  The main peak is produced by two-photon annihilation and the secondary
  peak (in the panel on the left) by soft-photon emission by final-state $e^\pm$ during $s$-channel scattering.
  For convenience, both input and output spectra are plotted in terms of an effective emission cross section.
  Fine structure above $\hbar\omega \sim m_ec^2$ in the highest-$p$ emission spectrum reflects the $\mu$-binning
in the MC code.  }
\end{figure*}\label{fig:spectra}

\subsection{Hard X-ray Spectrum:  Numerical Results}\label{s:spectrum}

The spectrum of photons escaping from a pair-loaded magnetic arcade is compared with the input spectrum
sourced by two-photon annihilation in Figure \ref{fig:spectra}.  For ease of comparison, the source spectrum
is plotted as the differential emission cross section multiplied by the photon energy, and the output spectrum is also converted
to an effective cross section.  The source spectrum (black curves) is 
strongly peaked near the kinematic limit $\hbar\omega_{\rm max} = 2\gamma m_ec^2$.
In general, in the hard X-ray band (as probed e.g. by RXTE,
NuStar and Suzaku measurements) the output spectrum closely aligns with the input spectrum.

The hard photons are mainly absorbed by photon collisions (compare the blue and black points in Figure \ref{fig:pairrates}).
At lower thermal momenta (left panels of Figure \ref{fig:spectra}),
the increased absorption seen at larger scattering depth is due to the increased photon source density and photon-photon opacity.
At higher thermal momenta (right panels), the output spectrum maintains a characteristic shape
around the peak at $\hbar\omega \gtrsim m_ec^2$ due to the effects of multiple scattering off $e^\pm$.

Scattering drives an upward flux of photons in frequency space.  This flux is uniform in
frequency when (i) the spectrum has the slope $dn_\gamma/d\ln\omega \sim$ constant (as is the case here below the peak) and (ii) the scattering
cross section is frequency-independent \citep{komp57}.  The source spectrum below the peak is preserved by scattering 
due to the vanishing divergence of the photon flux in frequency space, and also
because photons upscattered to $\hbar\omega \sim m_ec^2$ are converted back to pairs, instead of accumulating near the thermal peak.
(In detailed balance, the ratio of pairs to gamma rays near the peak
is $n_\pm/n_\gamma \sim (\sigma_{\gamma\gamma}/\sigma_{\rm ann})^{1/2} \sim B/\BQ$.)

\begin{figure}[t]
  \epsscale{1.2}
  \vskip -0.4in
\plotone{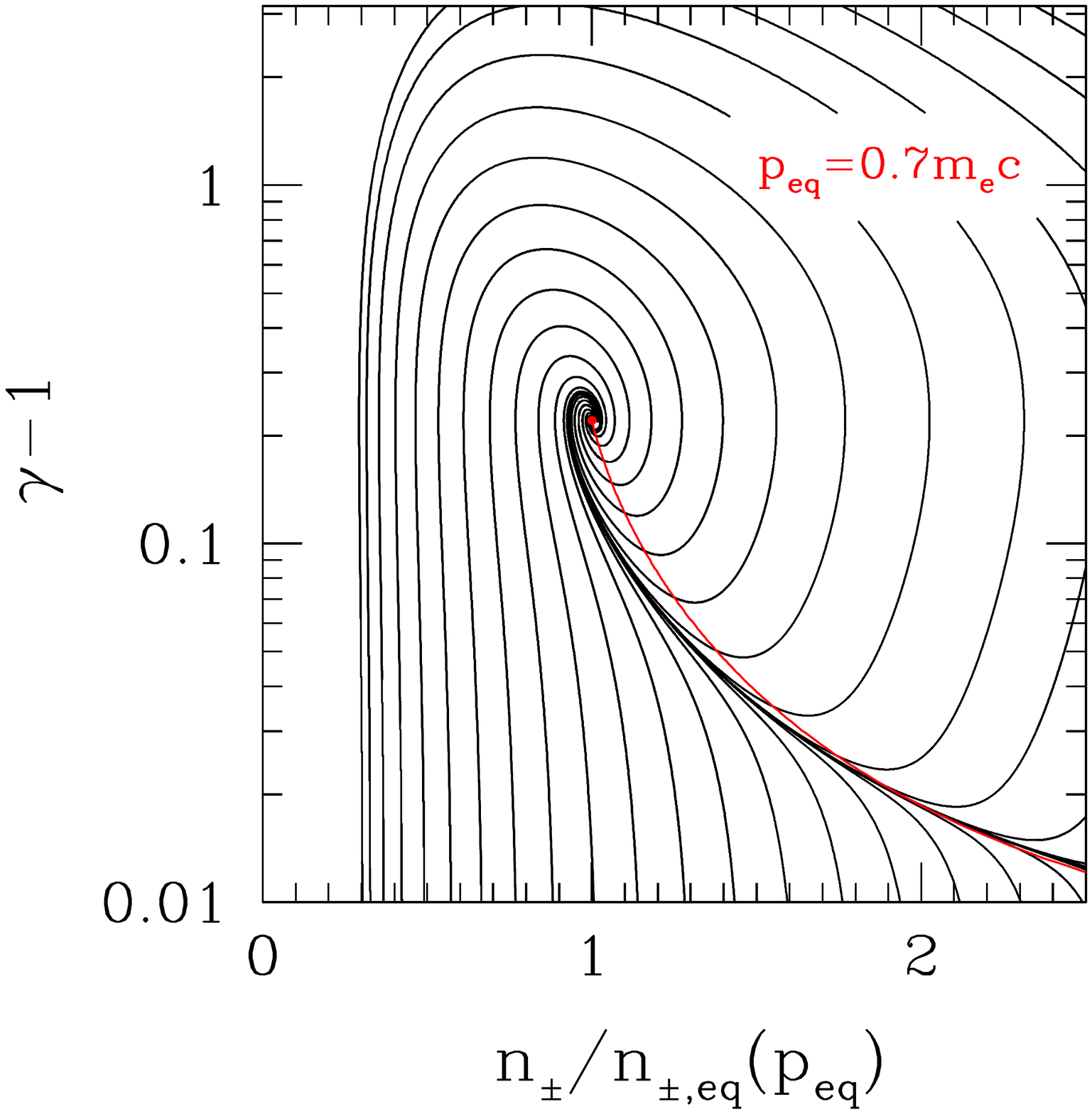}
\vskip -0.7in
\caption{Relaxation to combined energy and annihilation equilibrium (marked by the red dot) of a collisional, quasi-thermal $e^\pm$ plasma that is embedded in
  a sheared magnetic field stronger than $\sim 5\,\BQ$.  Black lines show the trajectory followed by the plasma.
  The flow is everywhere toward the equilibrium point.
  Here we take $p_{\rm eq} = 0.7\,m_ec$, $\alpha_n = 0.1$, $\hbar\omega_{\rm max} =
  m_ec^2$, and $\omega_{\rm max}/\omega_{\rm min} = 10^5$.  The red line corresponds to the energy equilibrium $\dot\gamma = 0$,
  as obtained by combining Equations (\ref{eq:ndot}) and (\ref{eq:udot}).}
\end{figure}\label{fig:attractor}

\subsection{Relaxation to Energy and Annihilation Equilibrium}\label{s:equil}

Now let us consider how a dense, quasi-thermal pair plasma relaxes to both energy and annihilation equilibrium within
a magnetic arcade carrying a strong current, $\tau_{\rm T,min} \gtrsim 1$.  
The MC simulations give the net rate of pair creation/annihilation, as a function of the $e^\pm$ thermal momentum
(Figure \ref{fig:pairrates}).   For a given scattering depth across the plasma, there is particular value of $p$ at which
creation balances annihilation.  Near this equilibrium momentum $p_{\rm eq}$, we Taylor expand,
\be\label{eq:ndot}
   {\dot n_\pm(p)\over \dot n_{\rm ann}} = \alpha_n\left({p\over p_{\rm eq}} - 1\right),
   \ee
   where $\dot n_{\rm ann}$ is given by Equation (\ref{eq:dotnann}).
The value of $p_{\rm eq}$ varies weakly with $N_\pm$, and $\alpha_n \sim 0.1$.  We also found that the analytic expression
(\ref{eq:taueq}) is a reasonable approximation to the pair density at which ohmic heating and radiative cooling are in
balance.  Then we can write
\be\label{eq:udot}
   {1\over \dot n_{\rm ann}}{d\over dt}(n_\pm \gamma) = -{\hbar\omega_{\rm max}/m_ec^2\over \ln(\omega_{\rm max}/\omega_{\rm min})}
   \left[1 - {n_{\pm,\rm eq}^2(p)\over n_\pm^2}\right].
\ee
Here the equilibrium pair density is a function of $p$ through Equation (\ref{eq:taueq}).

These equations are easy to integrate in the approximation where $p_{\rm eq}$ varies slowly with optical depth,
which is the case when the optical depth transverse to the confining magnetic field is  $\sigma_{\rm T} N_\pm = O(1)$.  Then
$n_{\pm,\rm eq}^2(p)/n_{\pm,\rm eq}^2(p_{\rm eq}) = (\beta^2\gamma/\beta_{\rm eq}^2\gamma_{\rm eq}) f(\beta_{\rm eq})/f(\beta)$
for a fixed value of $p_{\rm eq}$.  The result of the integration is shown in Figure \ref{fig:attractor}.

  When the plasma starts off warm and dense (top-right corner), it first cools off, until it reaches a low thermal
  energy where residual ohmic heating balances the strong radiative cooling ($\dot\gamma \simeq 0$).  Then,
  as the pairs annihilate, the plasma follows this
  quasi-equilibrium state (the red line) to the attractor point (red dot) where the plasma is in both annihilation and 
  energy equilibrium.  If the plasma starts off overdense and cool (bottom right), it heats up until it reaches the energy equilibrium
  line, after which the evolution is the same.  On the other hand, the energy equilibrium is not stable when the $e^\pm$ plasma
  is underdense.  If it starts off cool (bottom left), the plasma initially
  overheats, after which the pairs multiply to the point that radiative cooling sets in.  The trajectory is then the same
  as starting from the warm and overdense state.

\section{Infrared-Optical Emission}\label{s:optical}

The magnetar corona as described in this paper is a bright source of optical and near-IR photons,
emitting a flux much higher than expected from the Rayleigh-Jeans tail of the
surface $\sim$ keV blackbody.   The annihilation bremsstrahlung spectrum extends
from hard X-rays down to frequencies as low as $\sim 10^{14}$ Hz, where it is constrained
by (i) the plasma cutoff, (ii) self-absorption, and (iii)
induced electron scattering.  Optical-IR emission may also be strongly enhanced by plasma instabilities,
e.g. in the transition layer between the magnetosphere and the magnetar atmosphere (Section \ref{s:anom}).

\subsection{Plasma Cutoff}

The relatively high pair density considered here implies a high magnetospheric plasma frequency $\omega_{P\pm} = 2\pi\nu_{P\pm} 
= (4\pi n_\pm e^2/\gamma m_e)^{1/2}$ near the magnetar surface:
\ba\label{eq:nupe}
\nu_{P\pm}(R) &=&  \left[{2e^2 \tau_{\rm T}(R)\over \pi\gamma\sigma_{\rm T} m_e R}\right]^{1/2}\nn
&=& 5\times 10^{13}\,[\tau_{\rm T}(R)_1]^{1/2} \gamma^{-1/2} R_6^{-1/2}\quad{\rm Hz}.\nn
\ea
This frequency decreases outward approximately as
\be\label{eq:nupmr}
\nu_{P\pm}(r > R) = \nu_{P\pm}(R) \left[{B(r)\over B(R)}\right]^{1/2} \sim \nu_{P\pm}(R) \left({r\over R}\right)^{-3/2}.
\ee
Since low-frequency annihilation bremsstrahlung cuts off in magnetic fields weaker than $\sim 5\,\BQ$,
the plasma frequency remains above $10^{13}$ Hz in the emission zone, and emission at yet lower frequencies is
strongly suppressed.

\subsection{Inverse Annihilation Bremsstrahlung}

The process inverse to two-photon pair annihilation is two-photon pair creation.  A detailed account of
this self-absorption process is included in our MC procedure.  The absorption of soft photons requires target
photons near the threshold for single-photon pair conversion, which experience strong attenuation by
the enhanced $u$-channel resonant scattering off $e^\pm$.  We find (Figure \ref{fig:spectra}) that the escaping
low-frequency spectrum is very close to the source spectrum -- even when the pair plasma is optically thick
to scattering at lower frequencies.

The absorption process which must still be considered is the inverse of soft-photon emission during
$s$-channel $e^\pm$ backscattering (essentially, during single-photon pair annihilation).
We now write down an
expression for the angle-averaged absorption coefficient, which provides a reasonable measure of the attenuation
rate of soft photons in a curved magnetic field:
\be
\langle \alpha_\omega(s)\rangle_\mu = {\langle j_\omega(s)\rangle_\mu \over B_\omega(T)/2}.
\ee
We are only considering the emission of O-mode photons, hence the thermal intensity is
$B_\omega(T)/2 \simeq  T \omega^2/(2\pi)^3 c^2$.
Here, $4\pi \langle \omega j_\omega(s)\rangle_\mu$ is the volumetric rate of energy release in Equation (\ref{eq:udotann}).
In the spirit of the preceding calculations of photon emission, which assume a simple, monoenergetic $e^\pm$ distribution,
we take the non-relativistic limit and make the substitutions $f(\beta) \rightarrow 4\beta^2/3 \rightarrow 4T/3m_ec^2$, to get
\be
\langle \alpha_\omega(s)\rangle_\mu = {2\over 3}{\hbar\omega\over m_ec^2}\left({2\pi c\over\omega}\right)^3 {r_e^2 n_\pm^2 \over B/\BQ}.
\ee
Evaluating this at the top of a magnetic arcade ($r \sim 3R/2$), one gets
\be
\langle \alpha_\omega(s)\rangle_\mu\, r = 0.12 {[\sigma_{\rm T}n_\pm(r) r]^2
  \over \nu_{14}^2 B_{15} r_6}.
\ee
This is independent of temperature, in contrast with ordinary free-free absorption.  
Depending on the structure (width) of the magnetic loop, the plasma cutoff (Equation (\ref{eq:nupmr})) is
found to provide a marginally tighter constraint on low-frequency emission.

Soft photons are also absorbed during $t$-channel $e^\pm$ backscattering.   This process is 
equivalent to ordinary free-free absorption in a one-dimensional magnetized plasma.
Because the emission of soft photons during backscattering is proportional to the cross section
(see Appendix \ref{s:soft}),
\be
\langle j_\omega(t)\rangle_\mu  = {\sigma_{+-}(t)\over \sigma_{+-}(s)} \,\langle j_\omega(s)\rangle_\mu.
\ee
Recalling that $t$-channel backscattering has a relatively small cross section, we conclude that ordinary magnetic free-free absorption
can be ignored.

\subsection{Induced Electron Scattering}

Induced scattering limits the brightness of radiation escaping from a cloud of
quasi-thermal electrons \citep{zl69}.  When the scattering depth is of the order of unity,
the brightness temperature of the radiation cannot much exceed the kinetic temperature
of the scattering particles (here electrons and positrons).

A strong magnetization of
the pairs modifies the scattering cross section and the recoil energy loss.
A photon of energy $h\nu \ll m_ec^2$ scatters from a given state (with direction cosine $\mu=\cos\alpha$ and
occupation number $N(\nu,\mu)$) into a second state (with frequency $\nu_1 = \nu + \Delta\nu$,
direction cosine $\mu_1=\cos\alpha_1$ and
occupation number $N_1 = N(\nu,\mu_1)$) with a cross section $d\sigma_{\rm scatt}/d\mu_1 =
2\pi r_e^2\sin^2\alpha\sin^2\alpha_1$.  The recoil energy is
$h\Delta\nu = -{1\over 2}(\cos\alpha-\cos\alpha_1)^2 (h\nu)^2/m_ec^2$.  The contribution of
stimulated scattering to the time evolution of the occupation number is \citep{blandford76,sincell92}
\be
   {\partial N\over\partial t} + c(\hat k\cdot\bnabla)N = -2n_\pm cN
   \int d\mu_1  {d\sigma_{\rm scatt}\over d\mu_1} {\partial(\Delta\nu N_1)\over\partial\nu}.
\ee
Being interested in estimating the induced scattering rate within a cloud of scattering depth of
order unity, we substitute inside the integral the angle-averaged occupation number $\bar N$.
Performing the integral over $\mu_1$ gives
\ba\label{eq:indrate}
   &&{1\over N}{\partial N\over\partial t} + {c\over N}(\hat k\cdot\bnabla)N = \nn
   &&\quad\quad \sin^2\alpha\left(\cos^2\alpha + {1\over 5}\right) \sigma_{\rm T}n_\pm {h\over m_ec} 
   {\partial (\nu^2 \bar N)\over\partial \nu}.
\ea
   
The emission spectrum of annihilation bremsstrahlung corresponds to $\bar N \propto \nu^{-3}$, meaning
that the right-hand side of Equation (\ref{eq:indrate}) is negative and the high-intensity radiation is attenuated.  The
trigonometric factor takes a maximum value $6/25$;  hence, we estimate the attenuation coefficient as
\be\label{eq:alphaind}
\alpha_{\rm ind} \sim {\sigma_{\rm T}n_\pm\over 10} {T_b\over m_ec^2},
\ee
where the brightness temperature $T_b = h\nu \bar N$.

This limit on the escaping optical-IR flux can be applied to the magnetosphere, where
the scattering depth across the confining magnetic flux bundle is $\tau_{\rm T,\perp} \lesssim 1$;  and to 
the transition layer at the top of the magnetar atmosphere, which is much thinner, $\tau_{\rm T} \sim
10^{-2}\tau_{\rm T}(R)_1$.  In these two cases, the brightness temperature of the escaping low-frequency radiation
can exceed $m_ec^2$ by factors 10 and $10^3$, respectively.  This limiting flux is compared with the
optical-IR flux observed from some quiescent magnetars in Sections \ref{s:2259}, \ref{s:axp}.

\section{Sheared Magnetic Arcade:  \\ A Concrete Model}\label{s:arcade}

We now consider how the structure of a sheared magnetic field influences the emission by a trapped,
quasithermal pair plasma.   The specific example is a current-carrying arcade that is anchored
in a thin crustal shear zone, of width $\Delta l_\perp \sim 0.1\,R \sim 1$ km (Figure \ref{fig:dtwist}).  Similar
fault-like structures are seen to form in the global elastic-plastic-thermal simulations of \cite{TYO17}.
A large slippage along such a structure has been explored as a triggering model for a magnetar giant
flare \citep{TD01,parfrey13}, but here we are considering a quasi-static and small-amplitude twist.  The poloidal
magnetic field will be approximated as an axisymmetric dipole.  Deviations from
axisymmetry are readily found in the simulations of crustal yielding just described.  The single shear zone
discussed here could represent multiple shear zones of a smaller thickness.

We focus on field lines extending out to a distance $\sim 1.5\,R$, where the magnetic flux density
has dropped to $\sim 5\,\BQ$, corresponding to the threshold for annihilation bremsstrahlung emission.
Field lines closer to the magnetic pole could also support a plasma of similar density, in local
annihilation and energy equilibrium, but the emitting zone would not fill the arcade.

\begin{figure}[t]
  \vskip -0.5in
\epsscale{1.2}
\plotone{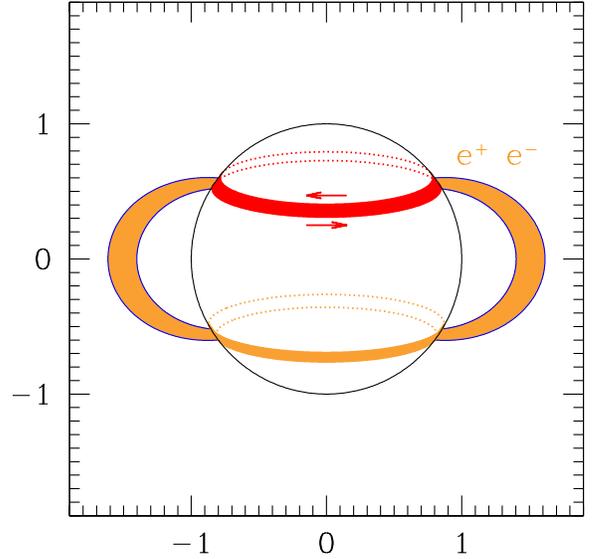}
\vskip -0.7in
\caption{Current-carrying magnetic arcade, here idealized as an axisymmetric structure filled with a transrelativistic
  and collisional $e^\pm$ gas, and anchored  at $\theta_s = {\rm asin}(\sqrt{2/3}) \pm 0.05$.  The magnetic field
  weakens from a surface strength of $10^{15}$ G to $\sim 5\,\BQ = 2\times 10^{14}$ G at the outer radius.
  Horizontal and vertical scales in units of the stellar radius.}
\end{figure}\label{fig:dtwist}

The crustal shear zone is concentrated between polar angles $\theta_s \pm \Delta l_\perp/2R$.
A field  line anchored
in the shear band extends to a maximum radius $R_{\rm max}(\theta_s) = R/\sin^2\theta_s$, where the flux density has
dropped to
\be
   B[R_{\rm max}(\theta_s)] = {(1-x_s^2/3)^3\over \sqrt{1+x_s^2}}\,B(R,\theta_s)
\ee
(here $x_s \equiv \sqrt{3}\cos\theta_s$).  The length of the field line is
\ba
l(\theta_s) &=&  2R_{\rm max}(\theta_s)\int_{\theta_s}^{\pi/2} \sin\theta (1+3\cos^2\theta)^{1/2}d\theta \nn
&=& {R_{\rm max}(\theta_s)\over\sqrt{3}}\left[\ln\left(x_s+\sqrt{1+x_s^2}\right) + x_s\sqrt{1+x_s^2}\right].\nn
\ea
The area of the (outer) magnetic flux surface is
\ba
A(\theta_s) &=& 2\int_{\theta_s}^{\pi/2} 2\pi r(\theta,\theta_s) {dl\over d\theta} d\theta \nn
&=& {\pi R_{\rm max}^2(\theta_s)\over 6\sqrt{3}}\biggl[13\ln\left(x_s + \sqrt{1+x_s^2}\right) \nn
&& \quad\quad\quad\quad\quad + \sqrt{1+x_s^2}\left(11x_s-2x_s^3\right)\biggr],\nn
\ea
where $r(\theta,\theta_s) = R_{\rm max} \sin^2\theta$.

Let us take, as an illustrative example, a surface magnetic field $B(R,\theta_s) = 10^{15}$ G and
a flux density about $1/4$ this at the top of the arcade.  This corresponds to $\sin^2\theta_s = 2/3$ and
\be\label{eq:quan1}
R_{\rm max} = {3\over 2}R; \quad l(\theta_s) = 1.99\,R; \quad A(\theta_s) = 16.5\,R^2.
\ee
For comparison, the area of the surface shear band and its hemispheric image is
\be\;\label{eq:quan2}
2A_\perp(R) = 4\pi \sin\theta_s R\Delta l_s = 1.03\,R^2\left({\Delta l_\perp\over 0.1~R}\right).
\ee

The scattering depth $\tau_{\rm T,\perp}$ transverse to the magnetic field is sensitive to the thickness of the crustal shear layer.
At any point on the flux bundle,
\be
   {\tau_{\rm T,\perp}(r)\over \tau_{\rm T}(R)}
     =  {2\Delta l_\perp \over R} \cdot {B_r(R,\theta_s)R\sin\theta_s\over B(R,\theta_s) r\sin\theta},
\ee
where $\tau_{\rm T}(R)$ is defined by Equation (\ref{eq:tauTR}) and we make use of the scaling
$n_\pm(r) \propto B(r)$.   Hence,
\be\label{eq:tauperp}
\tau_{\rm T,\perp}(R_{\rm max}) = 0.89\left({\Delta l_\perp\over 0.1~R}\right)\,\tau_{\rm T}(R)_1
\ee
for our default configuration.

\subsection{Radiative Output}

First consider the output in hard X-rays by annihilation bremsstrahlung.
Substituting for $l$ and $\Delta A_\perp$ into Equation (\ref{eq:edotvol2})
gives (for $\beta = 0.6$)
\ba\label{eq:edotann3}
   {d\dot E_{\rm ann}\over d\ln\omega} &=& 4\times 10^{34}\,{\tau_{\rm T}^2(R)_1 R_6\over B(R)_{15}}
   \left({\hbar\omega\over 100~{\rm keV}}\right)\nn
   && \times \left({\Delta l_\perp \over 1~{\rm km}}\right)\;\; {\rm erg~s^{-1}}.
\ea
The `hotspot' blackbody luminosity is, from Equation (\ref{eq:edotsurf2}),
\be\label{eq:edotbb3}
\dot E_{\rm bb} = 6\times 10^{34}\,\left({\varepsilon_{\rm bb}\over 0.25}\right)\tau_{\rm T}(R)_1R_6
\left({\Delta l_\perp \over 1~{\rm km}}\right)\;\; {\rm erg~s^{-1}}
\ee
and the effective temperature in one (ordinary) polarization mode is
\be
T_{\rm eff} = 0.65\,\left({\varepsilon_{\rm bb}\over 0.25}\right)^{1/2}
\left[{\tau_{\rm T}(R)_1\over R_6}\right]^{1/4}\quad{\rm keV}.
\ee

We can also compare the annihilation bremsstrahlung output in the near-IR to optical band
with the luminosity as limited by induced scattering.  Adapting Equation (\ref{eq:edotvol}) to $ \nu = 1.5-6\times 10^{14}$ Hz,
and substituting again for the length and cross section of the arcade gives
\be
   {d\dot E_{\rm ann} \over d\ln\nu} = 1.5\times 10^{29}\, {\nu_{14}\tau^2_{\rm T}(R)_1\over B(R)_{15}}
   \left({\Delta l_\perp\over 1~{\rm km}}\right)R_6^2\quad {\rm erg~s^{-1}}.
\ee
Whereas both $\dot E_{\rm ann}$ and $\dot E_{\rm bb}$ are proportional to the cross section of the arcade at the magnetar surface,
the bound on the optical-IR output from induced scattering is proportional to its surface area,
\ba\label{eq:limlum}
  {d\dot E_{\rm rad}\over d\ln\nu}\biggr|_{\rm max} &\sim& {c\over 2} 4\pi \left({\nu\over c}\right)^3 T_b A\nn
  &=& 0.9\times 10^{30}\,\nu_{14}^3\left({T_b\over 10\,m_ec^2}\right) R_6^2
  \quad {\rm erg~s^{-1}}.\nn
\ea

\subsection{Application to 1E 2259$+$586}\label{s:2259}

Here we make a limited comparison of these results with observations of the Anomalous X-ray Pulsar (AXP) 1E 2259$+$586, which
are available at both low and high frequencies.  We consider (i) the relative amplitude of the hard X-ray emission
  (at the highest X-ray energies measured, where the annihilation bremsstrahlung is expected to dominate)
  and the unabsorbed blackbody output; and (ii) the independent constraint from the fitted blackbody temperature.
The hard X-ray emission of 1E 2259$+$586 has a phase-averaged photon index
$-1.2$, corresponding to a slope of $+0.8$ for $dL_X/d\ln\omega$ \citep{vogel14}.  At photon energy 70 keV, the measured energy flux
reaches 30\% of the luminosity
of the blackbody component ($4\times 10^{34}$ erg s$^{-1}$ at a distance 3.2 kpc; \citealt{pizzocaro19}).
Matching these quantities to Equations (\ref{eq:edotann3}) and (\ref{eq:edotbb3}) for the annihilation bremsstrahlung and surface
blackbody emission implies
$\tau_{\rm T}(R) \simeq 8(\varepsilon_{\rm bb}/0.25)$ and $\Delta l_\perp \simeq 0.8 R_6^{-1} (\varepsilon_{\rm bb}/0.25)^{-1}$ km.

The blackbody temperature ($kT_{\rm bb} = 0.44$ keV; \citealt{pizzocaro19}) gives an independent measure of the radiative area.  If the
blackbody photons are radiated in only one polarization mode (here the O-mode), then $\varepsilon_{\rm bb} \sim 0.06$ 
  and the magnetospheric optical depth and width of the surface shear layer are inferred to be
$\tau_{\rm T}(R) \simeq 2$, $\Delta l_\perp \simeq 3.2\,R_6^{-1}$ km.
We require a reduction of at least a factor $\sim 4$ in the downward flux of positrons into the
magnetar atmosphere compared with the simplest kinematic estimate.  Ion-acoustic turbulence in the upper atmosphere
could have this effect (Section \ref{s:anom}).  The reduction factor could be larger, since part of the baseline blackbody
emission could be powered by internal dissipation and propagate in the extraordinary polarization mode.

The output in the optical-IR band is, in this model, directly tied to the output in hard X-rays, and
in quiescence is constrained to be $d\dot E_{\rm ann}/d\ln\nu = 7\times 10^{28}$ erg s$^{-1}$ at $\lambda = 2\,\mu{\rm m}$
(where the photon energy is only $6\times 10^{-6}$ of a 70 keV photon).  The measured ratio of IR to
thermal X-ray flux is $1.5\times 10^{-4}$, corresponding to $5\times 10^{-4}$ of the 70 keV output.  During the 2002 outburst of
1E 2259$+$586, this ratio was maintained as both flux components decayed \citep{tam04}.

We conclude that the directly emitted annihilation bremsstrahlung radiation cannot directly account for the measured optical-IR
output of 1E 2259$+$586, if the hard X-ray output matches observation.  The reprocessing of ultraviolet radiation into longer
wavelengths is an interesting alternative:  the output above 13.6 eV is inferred
to exceed $1.6\times 10^{30}$ erg s$^{-1}$, and so could power a significant part of the measured optical-IR flux.

Coherent plasma emission from the X-ray emitting flux bundles is also an interesting possibility here.
The plasma frequency in the inner magnetosphere and the atmospheric transition layer
is $\nu_P \gtrsim 5\times 10^{13}R_6^{-1/2}(\varepsilon_{\rm bb}/0.25)^{1/2}$ Hz (Equation (\ref{eq:nupe})).  This means that
O-mode photons formed by the coalecence of Langmuir waves can extend into the optical-IR band.

The dissipation in the atmospheric transition layer, as bounded by the kinetic energy flux of incident pairs, could  easily be high
enough to power the observed optical-IR emission.
Although the details of any coherent emission process are difficult to pin down, a considerable simplication arises here:  when
the intrinsic emission process is bright enough, the escaping flux depends mainly on the limiting effects of induced scattering.
Radiation escaping the magnetosphere has $T_b \lesssim 20\,m_ec^2$, given that the
scattering depth across the arcade is inferred to be $\tau_{\rm T,\perp}
\sim 0.7$ in 1E 2259$+$586 (Equation (\ref{eq:tauperp})), independent of the blackbody efficiency $\varepsilon_{\rm bb}$.
The output at $2\,\mu{\rm m}$, as limited by induced scattering off transrelativistic pairs,
is $\sim 6\times 10^{30}$ erg s$^{-1}$ (Equation (\ref{eq:limlum})), comparable to the measurement.
The limitation at higher, optical frequencies from induced scattering is
not significant, due to the $\sim \nu^3$ scaling.

Here we have not tried to match the surface magnetic field to the apparent spindown-determined dipole field
of 1E 2259$+$586, which is about $10^{14}$ as opposed to $10^{15}$ G.  This object has the longest spindown age of
any persistently bright magnetar, indeed longer than the age of the surrounding SNR.  The detection of X-ray emission
of comparable strength to other persistently bright AXPs
suggests the presence of stronger magnetic fields within the star, and has been interpreted as a signature of
a decay of the dipole field into higher multipoles \citep{TD96}.

\subsection{Other Anomalous X-ray Pulsars}\label{s:axp}

The source 4U 0142$+$61 gives a similar comparison with our model, with a hard X-ray photon index of $-1.0$,
ratio $3\times 10^{-4}$ of the $2.15\,\mu{\rm m}$ and 70 keV energy fluxes, and best-fit blackbody temperature of 0.46 keV \citep{hulleman04,tendulkar15}.
The case of 1E 1048.1$-$5937 \citep{yang16,archibald20} is more interesting, as it shows (i) strongly variable $20-70$ keV emission,
which fades below detectability in quiescence; and (ii) a smooth, sinusoidal, and weakly energy-dependent light curve, with some
structure emerging in the 10-20 keV band. The blackbody temperature also rises above $0.6$ keV during outbursts.  The IR flux is
less well sampled than the X-rays, but also shows coordinated changes and with a similar ratio of IR and X-ray energy fluxes
\citep{tam08}.

\subsection{Amplitude and Slope of the Hard X-ray Emission}

The persistent hard X-ray output of a magnetar is straightforwardly related to the area of its surface
covered by strongly sheared magnetic fields.  For example, SGR 1900$+$14 shows an energetically dominant hard X-ray
continuum with $60-70$ keV output an order of magnitude higher than the blackbody component \citep{gotz06,enoto17}.
This suggests a high current density and
scattering depth in the current-carrying arcade(s) ($\tau_{\rm T}(R) \sim 10^2$).  Our MC simulations (Section \ref{s:spectrum})
show that the annihilation bremsstrahlung source spectrum, $d^2n_\gamma/dtd\ln\omega \sim $ constant, is preserved by
$e^\pm$ scattering at such a large scattering depth.  Photons upscattered to an energy $\hbar\omega \sim m_ec^2$ do not accumulate into a thermal peak,
but instead are converted back to pairs, maintaining the high plasma density.

By contrast, SGR 1806$-$20 has a flatter X-ray spectrum that extends upward smoothly from $\sim$ keV photon energies,
and only shows a weak thermal peak \citep{mereghetti05b,enoto17}.  In this source,
we infer a reduced current density combined with a relatively large covering factor of the magnetar surface by the current-carrying arcades.
A full assessment of the output spectrum in this situation requires treating the interaction of magnetospheric
photons with the magnetar surface \citep{KT20}.

\section{Summary}\label{s:summary}

A quasi-thermal pair plasma embedded in a magnetic field stronger than $5\,\BQ \sim 2\times 10^{14}$ G is
a strong source of non-thermal X-ray photons, with a spectral slope similar to that observed
in the powerful hard X-ray emission of quiescent magnetars.  This QED process -- which
we have termed ``annihilation bremsstrahlung'' -- is 2-3 orders of magnitude brighter than classical bremsstrahlung.
Its detection points directly to the presence of super-QED magnetic fields.  The emission of photons 
in the 10-100 keV band is concentrated away from the surface of the magnetar (in contrast with
ordinary bremsstrahlung emission) but cuts off as the
magnetic field drops below $\sim 5\,\BQ$.    This process can be viewed as a soft-photon
correction to single-photon pair annihilation,
$e^+ + e^- \rightarrow \gamma \rightarrow e^+ + e^-$; part (but not all) of the emission is represented by
two-photon pair annihilation, with one photon rapidly reconverting back to a pair.
In weaker magnetic fields, the spectrum reverts to the annihilation line characteristic of pair
collisions in vacuo.

Our calculations employ a full MC realization of X-ray photon emission, transport, and
destruction, using the QED cross sections and rates derived by \cite{KT18,KT19}.
Careful attention is given to the emission and scattering of photons near the energy threshold for pair conversion;
these play a major role in energy transport through the plasma.  For example,
binning of the photon spectrum must be concentrated near
$\hbar\omega_{\rm max} \rightarrow m_ec^2/\sin\theta$, due to the strong $u$-channel resonance in the scattering
cross section.  A proper handling of photon collisions requires a gradual turn-on of the two-photon
pair creation process.

A combined treatment of annihilation bremsstrahlung and positron annihilation at the magnetar surface,
along with the reprocessing of the surface radiation by scattering, will be presented elsewhere \citep{KT20}.

Our results can be summarized as follows.

1. The persistent non-thermal emission of a magnetar is predicted to peak around $\hbar\omega \sim m_ec^2$.

2. The strength of the hard, rising X-ray continuum, relative to the surface blackbody, is proportional to the
optical depth through the magnetospheric pair plasma.  The measured fluxes of the thermal and non-thermal
components imply $\tau_{\rm T}(R) \gtrsim 10$.  

3. A transrelativistic and quasi-thermal pair plasma experiences strong collisional ohmic heating due to the enhanced scattering
cross section between electrons and positrons associated with the annihilation channel.

4. The competition between pair annihilation and two-photon pair creation regulates the thermal
momenta in the pair plasma to $p \sim (0.5-1)\,m_ec$: at higher temperatures, pair creation exceeds annihilation.

5. Photon emission balances ohmic heating when the $e^\pm$ plasma is $\sim 10-20$ times
denser than the minimum (charge-starvation) density that is needed to support the magnetospheric current.
In other words, the plasma is strongly in the collisional regime, and much less energetic than the collisionless double layer state
found by \cite{BT07}.

6. The state of combined energy and annihilation equilibrium is shown to be an attractor.

7. The currents that supply the rising, hard X-ray continuum of magnetars are strongly localized, as indicated
by the detection of thermal X-ray hotspots (e.g. \citealt{woods04,halpern05,bernardini09,an15}).  This is consistent
with the tendency of yielding in the magnetar crust to be strongly localized, either due to the action
of core magnetic stresses \citep{TYO17}, or global Hall drift \citep{gourg16}.  These currents need not be
sourced by the polar regions of the magnetar crust, and generally do not represent a shrinkage of a wider current toward
the polar cap, as suggested by \cite{B09}.

8. Intense ion-acoustic turbulence is excited in the transition layer between the magnetosphere and the thin
magnetar atmosphere.  The magnetospheric pairs feed back positively on the ion acoustic instability.
The current is imposed by the background magnetic shear, and cannot be reduced by scattering
of electrons or positrons off ion acoustic waves.  The downward flow of positrons into the atmosphere is reduced by
an order of magnitude if turbulent dissipation in the transition layer approaches the kinetic energy flux of
magnetospheric pairs.

9. The unusually bright optical-IR emission of magnetars offers a direct and complementary probe
of the same plasma that produces the hard X-rays.  The annihilation bremsstrahlung spectrum extends down to $\sim 10^{14}$ Hz,
below which it is suppressed by the plasma cutoff as well as annihilation bremsstrahlung absorption.
In the well-studied case of 1E 2259$+$686, which showed a coordinated decline in X-ray and optical-IR emission
following an outburst in 2002 \citep{tam04}, this process can directly contribute only a few percent of the measured optical-IR
flux.  The UV output is more than an order of magnitude brighter, and might power the optical-IR emission through reprocessing.

10. The surface plasma frequency is $\nu_{P\pm}(R)\gtrsim 10^{14}$ Hz, suggesting that
coherent emission by plasma turbulence could contribute to the optical-IR flux,
e.g. via Langmuir wave coalescence into photons of frequency $2\nu_{P\pm}$ \citep{eichler02}.  
We show that the measured fluxes are marginally consistent with the limitation due to
magnetic induced scattering in a plasma of transverse optical depth $\tau_{\rm T,\perp} \sim 1$.

11. Lower-frequency (radio to millimeter-wave) emission cannot be produced in the hard X-ray emitting plasma,
given the high plasma cutoff frequency.  The measured fluxes remain consistent with
plasma emission triggered by current-driven instabilities near the open-closed magnetic separatrix \citep{T08a,T08b},
where relativistic plasma flows are easier to sustain and the plasma density is lower.

Our work has further implications for plasma simulations of magnetar magnetospheres,
models of electrodynamic processes, and the mechanism of magnetic field decay, which we now describe.

\subsection{Comparison with Resonant $e^\pm$ Scattering Models for the Hard X-ray Continuum}

Resonant cyclotron scattering of soft, thermal X-rays by relativistic $e^\pm$ flowing beyond $10-30$ magnetar radii is a possible
source of the rising hard X-ray emission of magnetars \citep{FT07,BH07,B13a,wadiasingh18}.  Fine tuning is required
if this process is to explain energetically dominant $> 10$ keV emission, because the dissipation must
be localized in a relatively narrow flux bundle overlapping the polar cap.

It is natural for some
twist to be redistributed toward the polar cap by resistive evolution \citep{BT07, T08b, B09, CB17}, but
typically over much longer timescales than the initiation of an X-ray outburst.  Indeed, a delay
is observed in the increase in spindown torque relative to the triggering X-ray outburst \citep{archibald15,archibald20}.
This provides direct evidence that the peak X-ray emission is not powered by currents flowing near the polar cap.
Additional evidence in this direction comes from the behavior of 1E 2259$+$586, which is observed to emit hard X-ray photons
with at least 30\% the luminosity of the thermal X-ray bump, but has a spindown age much longer than any other active magnetar.

Another diagnostic is provided by the relative strength of the blackbody and rising hard X-ray spectral components.
In the detailed version of the resonant cyclotron scattering model developed by \cite{B13a,B13b}, the energy
carried by outflowing $e^\pm$ is comparable to that backflowing to the magnetar surface.  So one deduces
that $\dot E_{\rm bb}$ is comparable to the bolometric high-energy output.  In this model, the bolometric output is
concentrated above $\hbar\omega \sim m_ec^2$ and therefore exceeds the luminosity at 100 keV.
This implies a powerful blackbody component that is not seen in the persistent emission of the
most strongly bursting Soft Gamma Repeater (SGR) sources
(see \citealt{enoto17} for a broad compilation of magnetar X-ray spectra).

In particular, a dominance of the surface blackbody over the 10-100 keV X-ray continuum is expected
if the collisionless double layer described by \cite{BT07} is active in the polar regions of the magnetar
circuit.   In this solution to the magnetospheric charge flow, the inner parts of the circuit are marginally charge starved,
with $n_\pm \simeq n_{\pm,\rm min}$.  The $e^+$ and $e^-$ counterstream relativistically with approximately equal
kinetic energies.  

More generally, the strength of the blackbody spectral component depends on the degree of collisionality of
the magnetospheric pair plasma.  Inspecting Equations (\ref{eq:edotvol}) and (\ref{eq:edotsurf}) describing
the emission from a collisional and transrelativistic plasma, one sees that the ratio $\dot E_{\rm ann}/\dot E_{\rm bb}$ scales in proportion
to the optical depth.   As $\tau_{\rm T}(R)$ increases, there are relatively more annihilations in the magnetosphere
(rate $\propto n_\pm^2$) as compared with positrons impacting the surface (rate $\propto n_\pm$).
The most active SGRs are expected to have strong currents (and possibly finer structure in the crustal
shear pattern), leading to higher $\tau_{\rm T}(R)$ and a relatively stronger hard X-ray component

Finally, in all versions of the resonant scattering model, the observed spectrum
and flux depend strongly on orientation;  this has allowed the relative
orientation of the magnetic axis, rotation axis and line-of-sight to the observer to be tightly constrained
when fitting to the spectra of AXPs \citep{vogel14,an15}.   
As regards annihilation bremsstrahlung emission, we note that most of the emission occurs in a zone
which is marginally transparent to scattering.   Beaming effects are weaker but nonetheless present, because the
electrons and positrons move with a speed $\beta \sim 0.5-0.7$.
A detailed account is beyond the scope of this paper, as it must involve modelling of radiation transfer
through a curved magnetic field.

\subsection{Non-Local Sourcing of Pairs in the Magnetosphere}\label{s:nonlocal}

The emission of a hard continuum from localized parts of the inner magnetosphere, as proposed here, will have a profound
effect on the electrodynamics of neighboring zones carrying weaker currents, including the polar cap region.

First, the double layer solution in the inner magnetosphere \citep{BT07} depends on in situ pair creation by the direct
conversion of X-ray photons upscattered at the first Landau resonance,
\ba
X + e^\pm[n=0] &\;\rightarrow\;& e^\pm[n=1] \;\rightarrow\; \gamma + e^\pm[n=0] \nn
&\;\rightarrow\;& e^+ + e^- + e^\pm[n=0].
\ea
(This is kinematically possible in
magnetic fields stronger than $4\,\BQ$.)  The threshold energy for an $e^\pm$ to resonantly
scatter a X-ray of frequency $\omega_X$ is given by $\gamma_{\rm res} (1-\beta_{\rm res}\mu) = (B/\BQ) m_ec^2/\hbar\omega_X$,
and except for nearly collinear propagation scales inversely with $\omega_X$.  Thus, pair multiplication becomes
possible at much lower $\gamma_{\rm res}$ if the inner magnetosphere is bathed with X-rays much harder than
the surface blackbody.  Nonetheless, this additional source of pairs is also limited by rapid
electrostatic acceleration of the scattering charge through the resonant energy \citep{BT07}.

A second, and more powerful, effect involves the creation of pairs by collisions of gamma rays that are
emitted in the inner magnetosphere.  The flux of photons emitted by annihilation bremsstrahlung somewhat
above $\hbar\omega \sim m_ec^2$ can greatly exceed the flux of charges that is needed to support the magnetospheric
plasma.   Furthermore, the cross section for $\gamma + \gamma \rightarrow e^+ + e^-$ is enhanced
by a factor $B/\BQ$ in a super-QED magnetic field, and is further enhanced if one of the colliding photons
is much less energetic than the other \citep{KM86,KT18}.

The density of these gamma rays at points near the magnetar surface, outside the zones of peak current,
is $n_\gamma \sim L_\gamma/4\pi R^2 m_ec^3 \sim 3\times 10^{17}\,L_{\gamma,35}R_6^{-2}$ cm$^{-3}$.
The current density in a smoothly twisted dipole field, at an angle $\theta$ from the magnetic pole is,
for comparison, $J \simeq (cB/4\pi R)\sin^2\theta\Delta\phi_{\rm N-S}$,
where $\Delta\phi_{\rm N-S}$ is the net closed-field twist angle \citep{TLK02}.  The minimum
density of current-carrying pairs is, then, $n_{\pm,\rm min} = 1.7\times 10^{16}\,(\theta^2/\beta)\, B_{15}R_6^{-1}(\Delta\phi_{\rm N-S}/0.1)$
cm$^{-3}$ near the magnetic pole.
The cross section for the conversion of two equal-energy photons moving oppositely and in a direction perpendicular
to ${\bf B}$ is \citep{KT18}
\be
\sigma_{\gamma\gamma} = 6\sigma_{\rm T} \left({B\over \BQ}\right) \beta \left({m_ec^2\over \hbar\omega}\right)^6,
\ee
where $\beta$ is the speed of the created $e^\pm$ (before re-acceleration).  For example,
$\sigma_{\gamma\gamma} = 21\,B_{15}\,\sigma_{\rm T}$ when $\hbar\omega = 1.25\,m_ec^2$.  Then the optical
depth for collisions across a distance $\theta R$ (the cylindrical radius of a polar magnetic flux bundle) is
\be
\tau_{\gamma\gamma} \sim 1\,\theta B_{15}L_{\gamma,35} R_6^{-1}.
\ee
Photon collisions can easily supply mobile
charges in much of the closed magnetosphere ($\theta^2 \gtrsim 2\pi R/cP = 2\times 10^{-5}\, P_1^{-1}R_6$).
In a first estimate,
\be {\Delta n_\pm\over n_{\pm,\rm min}} \sim {n_\gamma \tau_{\gamma\gamma} \over n_{\pm,\rm min}}
\sim 20\,\theta^{-1} {L_{\gamma, 35}^2 \over R_6^2 (\Delta\phi_{\rm N-S}/0.1)}.
\ee

The composition of the current flowing at intermediate radius through the closed magnetosphere is of interest
for other reasons.  These charges will rescatter surface blackbody photons into a declining power-law tail
in the 1-10 keV band \citep{FT07,nobili08a}.  The voltage sustained in this zone could also be significantly
reduced by non-local sourcing of pairs, and the ohmic timescale lengthened.

\subsection{Implications for Current Decay}

Identifying the process responsible for the hard X-ray emission of magnetars allows some interesting inferences
to be drawn about the origin of the transient behavior frequently observed.
The ohmic decay time of the magnetospheric current supported by a transrelativistic, collisional pair gas is,
from Equations (\ref{eq:sigpm2}) and (\ref{eq:sigmacond}),
\ba\label{eq:tohm}
t_{\rm ohm} &=& {4\pi \sigma_{\rm cond} l_{\rm shear}^2\over c^2} \nn
&=& {4\alphem B/\BQ\over \pi \gamma} {l_{\rm shear}^2\over cr_e} = 8\times 10^3{B_{15}l_{\rm shear,5}^2\over\gamma}\quad {\rm yr}.\nn
\ea
This is comparable to the magnetar lifetime if the magnetic field is smoothly sheared over a scale $l_{\rm shear} \sim 1$ km,
but is correspondingly shorter if the deformation pattern in the crust is more irregular.

The persistent X-ray (and IR) emission of magnetars is typically observed to decay over weeks to months following an outburst.
This behavior is seen both following magnetar giant flares (e.g. \citealt{woods01}) and following lower-energy
outbursts, especially in AXPs (e.g. \citealt{ibrahim01,woods04,tam04,halpern08,rea09,yang16,archibald20}).  There is an important distinction
between the outbursts of SGRs and AXPs:   in the first case, the energy of the bright, short-duration X-ray
transient can exceed the energy radiated in the subsequent afterglow by a factor $\sim 100$, whereas for AXPs a short-duration transient
associated with the brightening is usually energetically subdominant, if detected at all.   In the latter case, a process injecting
twist into the external magnetic field at the beginning of the outburst would have to be electromagnetically inefficient.

Various mechanisms have been proposed to produce such transient, decaying X-ray emission:   conductive heating of
the magnetar surface by a trapped fireball during a giant flare \citep{TD95};  volumetric heating of the upper
crust during an outburst, driven either by ohmic heating \citep{lyub02} or by plastic damping of Alfv\'en waves
\citep{LB15};  surface heating by bombarding charges driven by excited magnetospheric currents \citep{T00,B09},
which may be sustained by persistent plastic flow in crustal shear zones following a rapid outburst, as seen in the ab initio yielding calculations
of \cite{TYO17}.  

The ohmic timescale (\ref{eq:tohm}) is too long to represent the observed flux decay if the magnetic field is smoothly sheared.
When the magnetic field is sheared on a small scale, the decay profile must represent the continued rate of forcing of
magnetospheric currents by crustal deformations.  Indeed, the yield calculations of \cite{TYO17} show continued creep
along fault-like structures:  the energy release decays overall as a power-law in time, but can be interrupted by spasmodic aftershocks.

\acknowledgements
This research was supported by an NSERC Discovery grant.

\appendix

\section{Soft Photon Emission During Electron-Positron Annihilations}\label{s:soft}

Here we demonstrate the equivalence between soft-photon emission during $e^+$-$e^-$ backscattering, and the
1-$\gamma$/1-p channel of two-photon pair annihilation.   The pair is assumed to be sufficiently
relativistic ($\beta > 0.24\,B_{15}^{1/3}$ in the center-of-momentum frame) that scattering is
dominated by the annihilation channel ($s$ channel), as depicted in Figure \ref{fig:feynman}.   The emission
of one hard, pair-converting photon in combination with one soft photon may, in other words, be viewed as
a soft-photon correction to the single-photon annihilation process, $e^+ + e^- \rightarrow \gamma
\rightarrow e^+ + e^-$,
\be
e^+ + e^- \rightarrow \gamma + \gamma_{\rm soft} \rightarrow e^+ + e^- + \gamma_{\rm soft}.
\ee
Here the soft-photon line is attached to one of the charged particles in the initial state.  But it
may also be attached to one of the final-state $e^\pm$ (Figure \ref{fig:feynman2}),
\be
e^+ + e^- \rightarrow \gamma \rightarrow e^+ + e^- + \gamma_{\rm soft}.
\ee
These two channels contribute equally to the soft-photon output.  Hence, when the $s$-channel dominates
$e^+$-$e^-$ backscattering, the net cross section for soft photon emission should be twice the two-photon annihilation
cross section in the 1-$\gamma$/1-p channel.  (No such enhancement is present in the annihilation
channels that produce two photons above or below the pair conversion threshold.)

\begin{figure}
\epsscale{0.7}
\plotone{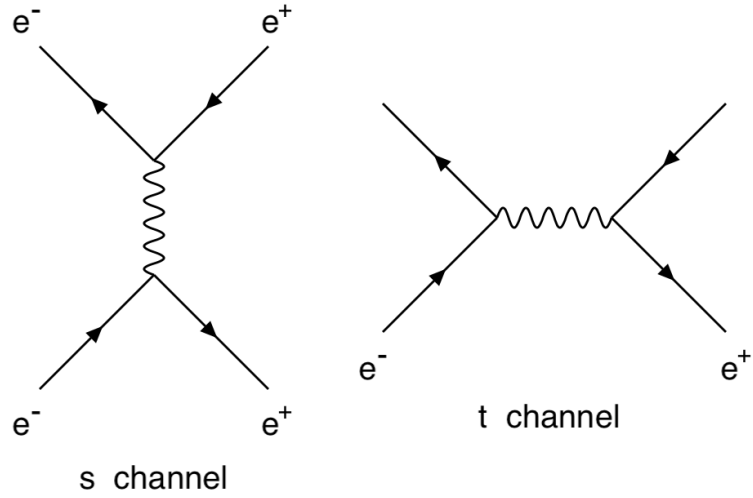}
\vskip -0.1in
\caption{Contributions to the matrix element for $e^+$-$e^-$ backscattering.
  Left panel:  annihilation channel ($s$ channel).  Right panel:  scattering channel ($t$ channel).
  In the presence of a strong magnetic field, the pair annihilates into a real photon and there is no
  interference between the two channels, i.e., there is a vanishing contribution to the cross section
  from the cross term between the two  diagrams \citep{KT19}.}
\end{figure}\label{fig:feynman}

The soft-photon emission cross-section can be directly related to the $e^+$-$e^-$ backscattering cross section
via \citep{LLT4}
\be\label{eq:soft}
{\omega\over 2\pi}{d^2\sigma_{\rm br}\over d\mu d\omega} = 4\alphem\left({\omega\over 2\pi c}\right)^2
\left({\varepsilon\cdot p_f\over k\cdot p_f} - {\varepsilon\cdot p_i\over k\cdot p_i}\right)^2\,\sigma_{+-},
\ee
where $k^\mu$ and $\varepsilon^\mu$ are the polarization 4-vectors of the soft photon, and $p_i^\mu \simeq -p_f^\mu$
can be chosen as the initial and final 4-momenta of the backscattering electron, restricted to the lowest
Landau state.  The factor 4 is the enhancement
in the soft-photon factor compared with that representing the emission of a single soft photon by
a single electron line.  (There is a factor 2 enhancement in the matrix element due to the presence of
distinct electron and positron lines.)  Setting $p^z_i = -p^z_f = \gamma\beta m_ec$ and $E_f = E_i = \gamma m_ec^2$,
we have
\be\label{eq:sigbr}
{\omega\over 2\pi}{d^2\sigma_{\rm br}\over d\mu d\omega} = {4\alphem\over \pi^2} {\beta^2(1-\mu^2)\over (1-\beta^2\mu^2)^2}\sigma_{+-}
\ee
for emission of an O-mode photon with direction cosine $\mu$ and $\varepsilon^z = (1-\mu^2)^{1/2}$.

\begin{figure}
\epsscale{0.8}
\plotone{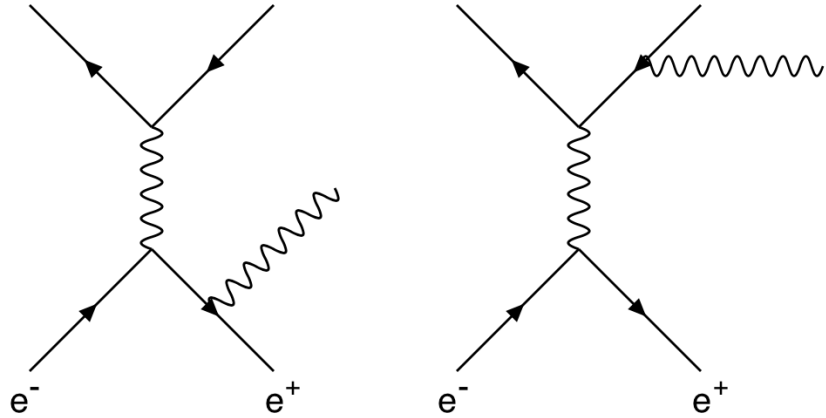}
\vskip -0.4in
\caption{Left panel:  a soft-photon line that is attached to the initial state $e^+$ (or $e^-$) represents
  a two-photon decay of the pair.  However, this diagram and its twin (with the soft photon line attached
  to the initial $e^-$) does not exhaust the possibilities for soft-photon emission associated with
  $e^+$-$e^-$ backscattering through the annihilation channel.   Right panel: the soft-photon
  line can also be attached to one of the final-state particles.  Hence the soft-photon emission cross section
  is a factor 2 larger than that implied by the two-photon decay cross section.}
\end{figure}\label{fig:feynman2}

The $e^\pm$ backscattering cross section (Figure \ref{fig:sigma_pp}) can be written as a discrete sum over the $s$ and $t$ channels
\citep{KT19}
\be\label{eq:sigpm}
\sigma_{+-} = \sigma_{+-}(t) + \sigma_{+-}(s) = {\pi r_e^2\over 4\beta^4\gamma^6}\left[ 1 + 2\pi\gamma \beta^2
  {m_ec^2\over \hbar\Gamma_{\gamma\rightarrow e^++e^-}}\right],
\ee
where \citep{KT18}
\be\label{eq:gampm}
{\hbar \Gamma_{\gamma\rightarrow e^++e^-}\over m_ec^2} = {\alphem B/\BQ\over 2\beta\gamma^3}e^{-2\gamma^2\BQ/B}
\ee
is the width of a pair-converting photon with $\mu = 0$ and energy $\hbar\omega \leq m_ec^2[1 + (1+2B/\BQ)^{1/2}]$.
Equation (\ref{eq:gampm}) breaks down when the photon is energetic enough for one of the created
$e^\pm$ to reside in the first excited Landau state, hence the upper bound on $\omega$.
The exponential factor in $\Gamma_{\gamma\rightarrow e^++e^-}$, which is close to unity for the annihilation of
a transrelativistic pair in a super-QED magnetic field,
arises from the vertex factor described following Equation (\ref{eq:sigma_ann_m1m2}).
In this case the photon is just above threshold for pair conversion, $\hbar\omega\sin\theta \simeq 2m_ec^2$.
There are two similar vertices in the $s$-channel contribution to  $\sigma_{+-}$, representing the annihilation
and regeneration of the pair.  Restoring the relevant vertex factors to $\sigma_{+-}(s)$ (which were set to unity
by \citealt{KT19}) implies multiplying by $e^{-4\gamma^2B_Q/B}$, giving the right term of Equation (\ref{eq:sigpm2}).

Hence the soft-photon cross section due to $s$-channel backscattering is
\be\label{eq:sigbr2}
\omega{d^2\sigma_{\rm br}\over d\mu d\omega} = {8\pi r_e^2\over B/\BQ} {\beta\over\gamma^2}  {(1-\mu^2)\over (1-\beta^2\mu^2)^2}
e^{-2\gamma^2\BQ/B}.
\ee
By way of comparison, the two-photon annihilation cross section (\ref{eq:sigma_ann_m1m2}) can be re-written as
\be\label{eq:sigann1}
\omega_1 {d^2\sigma_{\rm ann}\over d\mu_1 d\omega_1} = 2|\mu_2|{d^2\sigma_{\rm ann}\over d\mu_1d|\mu_2|}
= {4\pi r_e^2\over B/\BQ} {\beta\over\gamma^2}  {(1-\mu_1^2)\over (1-\beta^2\mu_1^2)^2}e^{-2\gamma^2\BQ/B}.
\ee
in the regime where $\omega_1 \ll \omega_2$ and $|\mu_2| = (\omega_1/\omega_2) \mu_1 \ll \mu_1$.  The factor 2 in the
middle term arises from the choice of photon 1 or 2 as the soft photon.  For consistency, we have also restored the
exponential vertex factor that was also set to unity in Equation (\ref{eq:sigma_ann_m1m2}).  One sees that Equation (\ref{eq:sigann1})
falls short by a factor ${1\over 2}$ in comparison with the soft-photon prediction (\ref{eq:sigbr2}),
for the reason described in the first paragraph of this Appendix.

\end{document}